\documentclass[12pt,headings=big,numbers=noenddot,DIV=14,a4paper]{scrartcl}%

\pdfoutput=1

\usepackage{amsmath,amssymb,amsfonts,soul}
\usepackage[normal,font=small,labelfont=bf,labelsep=period]{caption}
\usepackage[pdftex]{color,graphicx}
\usepackage{subfigure}
\usepackage[english]{babel}
\usepackage[compress]{cite}
\addtokomafont{disposition}{\rmfamily\boldmath}

\usepackage[dvipsnames]{xcolor}
\definecolor{darkblue}{rgb}{0,0.2,0.6}
\usepackage[linktoc=page,bookmarks=false,colorlinks=false,
linkbordercolor=RoyalBlue,citebordercolor=ForestGreen,
urlbordercolor=CornflowerBlue]{hyperref}

\numberwithin{equation}{section}
\newcommand{\abs}[1]{|#1|} 

\DeclareFontFamily{OT1}{pzc}{}
\DeclareFontShape{OT1}{pzc}{m}{it}{<-> s * [1.10] pzcmi7t}{}
\DeclareMathAlphabet{\mathpzc}{OT1}{pzc}{m}{it}
\DeclareMathOperator{\tr}{tr} 

\title{Exploring the Higgs sector\\ of a most natural NMSSM}
\author{\normalsize Riccardo
Barbieri$^a$, Dario Buttazzo$^a$, Kristjan Kannike$^{a,b}$, Filippo Sala$^a$, Andrea Tesi$^a$}
\date{\normalsize{\it  $^a$Scuola Normale Superiore and INFN, Piazza dei Cavalieri 7, 56126 Pisa, Italy\\
$^b$National Institute of Chemical Physics and Biophysics, R\"avala 10, Tallinn, Estonia}}

\begin{document}
\begin{titlepage}

\maketitle
\thispagestyle{empty}

\begin{abstract}
\centerline{\bf Abstract}\medskip
\noindent
The Next to Minimal Supersymmetric Standard Model (NMSSM) with a Higgs-singlet coupling $\lambda$ close to unity, moderate $\tan{\beta}$ and stop masses below 1 TeV minimizes the fine tuning of the electroweak scale, making possible that the lightest new particles, except perhaps for the LSP, be part of the extended Higgs system. We show how the measurements of the couplings of the 126 GeV Higgs boson constrain the region of the physical parameters of a generic NMSSM most relevant to this context. In the same region of parameter space we determine the cross section for the production of a heavier CP-even scalar together with its total width and its most relevant branching ratios. For comparison we show the same analysis for the MSSM. We also comment on how a coupling $\lambda \gtrsim 1$ can be compatible with gauge coupling unification.
\end{abstract}

\vfill
\noindent\line(1,0){188}\\\medskip
\footnotesize{E-mail: \scriptsize\tt{\href{mailto:barbieri@sns.it}{barbieri@sns.it}, \href{mailto:dario.buttazzo@sns.it}{dario.buttazzo@sns.it}, \href{mailto:kannike@cern.ch}{kannike@cern.ch}, \href{mailto:filippo.sala@sns.it}{filippo.sala@sns.it}, \href{mailto:andrea.tesi@sns.it}{andrea.tesi@sns.it}}}

\end{titlepage}

\section{Introduction}
\label{sec1}

The lack of signals so far from the direct production of supersymmetric particles calls for a reconsideration of the strategy to search for supersymmetry. Focussing on naturalness, which to us still looks to be the best motivated guideline, 
special attention is paid to the s-particles that have the largest influence on the quadratic terms in the Higgs potential, namely the two stops, the left-handed bottom, the higgsinos and, indirectly through the stops,  the gluino \cite{Dimopoulos:1995mi,Cohen:1996vb,Barbieri:2009ev,Papucci:2011wy}.

The mounting evidence for a Higgs boson $h_{\text{LHC}}$ at 126 GeV with Standard-Model-like properties \cite{Aad:2012tfa,Chatrchyan:2012ufa,ATLAS,CMS,tevatron:2013} also has a significant impact on these issues.
While the identification of the new resonance with the lightest Higgs boson of a supersymmetric model is a well-motivated possibility, the measured mass is in serious tension with maintaining naturalness in the Minimal Supersymmetric Standard Model (MSSM), requiring either large stop masses, above a TeV, in the case of negligible mixing, or a large trilinear $A_t$ term. At the heart of this problem is the fact that the quartic terms of the Higgs potential in the MSSM are controlled by the weak gauge couplings. 

The situation is different in the Next-to-Minimal Supersymmetric Standard Model (NMSSM), where a singlet superfield $S$ couples to the Higgs superfields, $H_u$ and $H_d$, via the Yukawa-like coupling $\lambda S H_u H_d$ \cite{Fayet:1974pd} (see \cite{Ellwanger:2009dp} for a review).  
On the one hand, the inclusion of this coupling allows to get a 126 GeV Higgs boson mass with both the stop masses well below a TeV. On the other hand, values of $\lambda \gtrsim 1$ suppress the sensitivity of the Higgs vacuum expectation value (VEV) with respect to changes in the soft supersymmetry-breaking masses, thus still keeping the fine tuning at a moderate level even for stop masses up to 1 TeV \cite{Barbieri:2006bg,Hall:2011aa,Agashe:2012zq}.%
\footnote{A recent analysis \cite{Gherghetta:2012gb} finds that the fine tuning in the NMSSM with a scale-invariant superpotential can be above $5\%$ for stop masses up to $1.2$ TeV and gluino masses up to 3 TeV for $\lambda \approx 1$, moderate $\tan{\beta}$ and the messenger scale at 20 TeV.}
We shall comment below on how a coupling $\lambda \gtrsim 1$ can be made compatible with gauge coupling unification.

In general terms, to see whether the newly found resonance at 126 GeV is part of an extended Higgs system is a primary task of the current and future experimental studies. Given the above motivations, this appears to be especially true  for the extra Higgs states of the NMSSM, which might be the lightest new particles of a suitable supersymmetric model, except perhaps for the lightest supersymmetric particle (LSP).
A particularly important question is how the measurements of the couplings of $h_{\text{LHC}}$, current and foreseen, bear on this issue, especially in comparison with the potential of the direct searches of new Higgs states.\footnote{For recent studies see e.g. Refs. \cite{Gupta:2012fy,DAgnolo:2012mj}.}

Not the least difficulty that one encounters in attacking these problems is the number of parameters that enter the Higgs system of the NMSSM, especially if one does not want to stick to a particular version of it but rather wishes to consider the general case.
Here we aim at an analytic understanding of the properties of the Higgs system of the general NMSSM, trying to keep under control as much as possible the complications due to the proliferation of model parameters and avoid the use of benchmark points.

The content of the paper is the following. In Section 2 we establish some relations between the physical parameters of the CP-even Higgs system valid in the general NMSSM. In Section 3 and 4 we consider two limiting cases in which one of the CP-even scalars is decoupled, determining  in each situation the sensitivity of the measurements of the couplings of $h_{\text{LHC}}$, current and foreseen, as well as the production cross sections and the branching ratios (BR) for the new intermediate scalar. In Section 5 we compare one of these NMSSM cases with the much studied MSSM, using as much as possible the same language. In Section 6 we illustrate a possible simple and generic extension of the NMSSM that can make it compatible with standard gauge unification even for a coupling $\lambda \gtrsim 1$. Section 7 contains a summary and our conclusions.

\section{Physical parameters of the CP-even Higgs system in the general NMSSM}
\label{sec2}

Assuming a negligibly small violation of CP in the Higgs sector, we take as a starting point the form of the squared mass matrix of the neutral CP-even Higgs system in the general NMSSM: 
\begin{equation}\label{scalar_mass_matrix}
{\cal M}^2=\left(
\begin{array}{ccc}
m_Z^2 \cos^2\beta+m_A^2 \sin^2\beta & \left(2 v^2 \lambda ^2-m_A^2-m_Z^2\right) \cos\beta \sin\beta &  v M_1  \\
 \left(2 v^2 \lambda ^2-m_A^2-m_Z^2\right) \cos\beta \sin\beta & m_A^2 \cos^2\beta+m_Z^2 \sin^2\beta +\delta_t^2 &  v M_2  \\
  v  M_1 &  v M_2 & M_3^2
\end{array}
\right)
\end{equation}
in the basis $\mathcal{H} = (H_d^0, H_u^0, S)^T$. In this equation 

\begin{equation}
\label{mHcharged}
m_A^2 = m_{H^{\pm}}^2 - m_W^2 +\lambda^2 v^2,
\end{equation}
where $m_{H^{\pm}}$ is the physical mass of the single charged Higgs boson, $v \simeq 174$ GeV, and
\begin{equation}
\delta_t^2 = \Delta_t^2(m_{\tilde t_1}, m_{\tilde t_2}, \theta_{\tilde t})/ \sin^2\beta
\label{delta-t}
\end{equation}
is the well-known effect of the top-stop loop corrections to the quartic coupling of $H_u$, with $m_{\tilde t_{1,2}}$ and $\theta_{\tilde t}$ physical stop masses and mixing.
We neglect the analogous correction to Eq. \eqref{mHcharged} \cite{Djouadi:2005gj}, which lowers $m_{H^{\pm}}$ by less than 3 GeV for stop masses below 1 TeV. 
We leave unspecified the other parameters in Eq. (\ref{scalar_mass_matrix}), $M_1, M_2, M_3$, which are not directly related to physical masses and  depend on the particular NMSSM under consideration.

The vector of the three physical mass eigenstates  $\mathcal{H}_{\rm ph}$ is related to the original scalar fields by
\begin{equation}
\mathcal{H} = R^{12}_{\alpha} R^{23}_{\gamma} R^{13}_{\sigma} \mathcal{H}_{\rm ph} \equiv R \mathcal{H}_{\rm ph},
\label{rotation_matrix}
\end{equation}
where $R^{ij}_\theta$ is the rotation matrix in the $i j$ sector by the angle $\theta = \alpha,\gamma,\sigma$.

Defining $ \mathcal{H}_{\rm ph} = (h_3, h_1, h_2)^T$, we have
\begin{equation}
R^T {\cal M}^2 R = \mathrm{diag}(m_{h_3}^2, m_{h_1}^2, m_{h_2}^2).
\label{diag_matrix}
\end{equation}
We identify $h_1$ with the state found at the LHC, so that $m_{h_1} = 125.7$ GeV. For simplicity we shall always consider 
$h_1$ as the lightest CP-even state, although other cases may be compatible with current data.%
 \footnote{Our analysis can be applied as well to the case of a new state below $h_{\text{LHC}}$. See \cite{King:2012tr} and references therein for a recent study of the NMSSM that especially emphasizes this case.}
From Eq. (\ref{rotation_matrix}) $h_1$ is related to the original fields by
\begin{equation}
h_1 = c_{\gamma} (-s_{\alpha} H_d + c_\alpha H_u) + s_{\gamma} S,
\end{equation}
where $s_\theta = \sin{\theta}, c_\theta = \cos{\theta}$. Similar relations, also involving the angle $\sigma$, hold for $h_2$ and $h_3$.

These angles determine  the couplings of $h_1 = h_{\text{LHC}}$  to the fermions or to vector boson pairs, $VV = WW, ZZ$, normalized to the corresponding couplings of the SM Higgs boson. Defining  $\delta = \alpha - \beta +\pi/2$, they are given by (see also \cite{Cheung:2013bn,Choi:2012he})
\begin{equation}
\frac{g_{h_1tt}}{g^{\text{SM}}_{htt}}= c_\gamma(c_\delta +\frac{s_\delta}{\tan\beta}),~~\frac{g_{h_1bb}}{g^{\text{SM}}_{hbb}}= c_\gamma(c_\delta -s_\delta \tan\beta ),~~\frac{g_{h_1VV}}{g^{\text{SM}}_{hVV}}=  c_\gamma c_\delta.
\label{h1couplings}
\end{equation}
\begin{figure}
\begin{center}
\includegraphics[width=.48\textwidth]{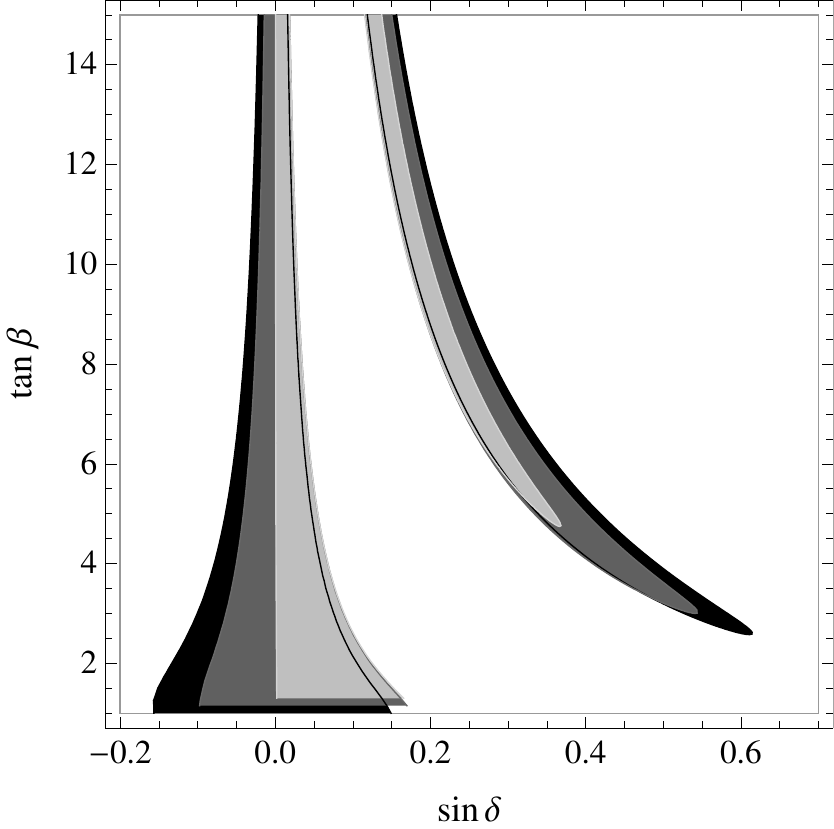}\hfill
\includegraphics[width=.48\textwidth]{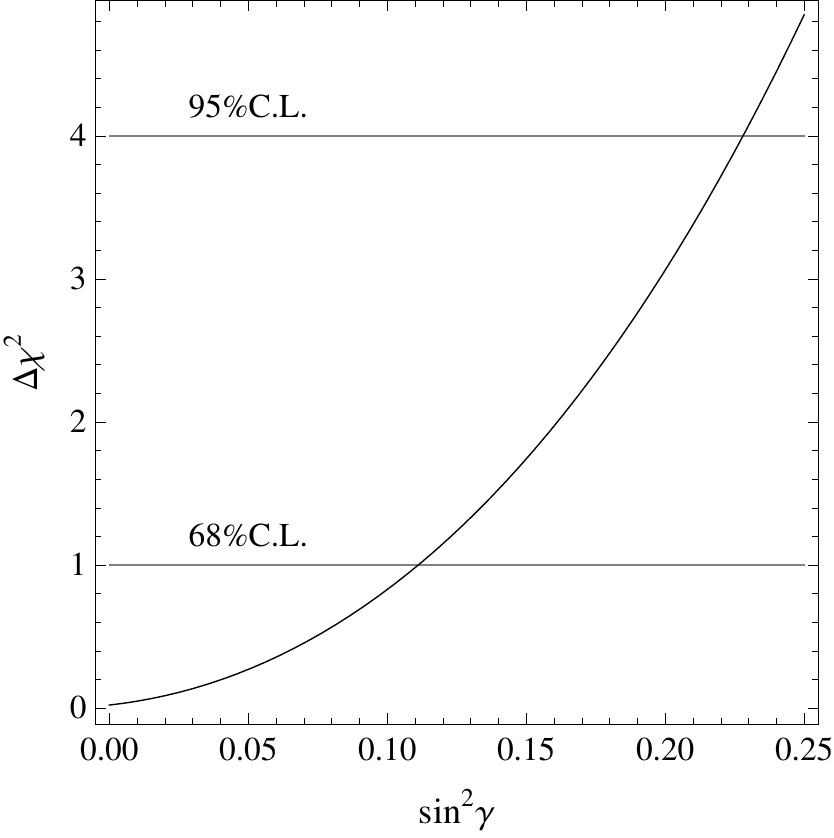}
\caption{\label{fig:FIT}\small Fit of the measured signal strengths of $h_1 = h_{\text{LHC}}$. Left: 3-parameter fit of $\tan \beta$, $s_{\delta}$ and $s_{\gamma}^{2}$. The allowed regions at 95\%C.L. are given for $s_{\gamma}^{2} = 0$ (black), $0.15$ (dark grey), and $0.3$ (light grey). The regions overlap in part, but their borders are also shown. Right: Fit of $s_{\gamma}^{2}$ in the case of $\delta = 0$.
}
\end{center}
\end{figure}
The fit of all ATLAS \cite{ATLAS}, CMS \cite{CMS} and TeVatron \cite{tevatron:2013} data collected so far on the various signal strengths of $h_{\text{LHC}}$ gives the bounds on $\delta$ for different fixed values of $\gamma$ shown in the left of Figure~\ref{fig:FIT} and the bound on $\gamma$ for $\delta= 0$ shown in the right of Figure~\ref{fig:FIT}. To make this fit, we adapt the code provided by the authors of \cite{Giardino:2013bma}. As stated below, we do not include in this fit any supersymmetric loop effects. Note that in the region of $s_\delta$ close to zero, a larger $s_\gamma^2$ forces $\delta$ to take a larger central value.

The matrix equation (\ref{diag_matrix}) restricted to the $1 2$ sector gives three relations between the mixing angles $\alpha, \gamma, \sigma$ and the physical masses $m_{h_1,h_2,h_3}, m_{H^{\pm}}$ for any given value of $\lambda, \tan{\beta}$ and $\Delta_t$. In terms of the $2\times 2$ submatrix $M^2$ in the $1 2$ sector of ${\cal M}^2$, Eq. (\ref{scalar_mass_matrix}),
 these relations can be made explicit as
\begin{align}
  s_\gamma^{2} &=  \frac{ \det M^{2} + m_{h_1}^{2} (m_{h_1}^{2} - \tr M^{2})}{(m_{h_1}^{2} - m_{h_2}^{2}) 
  (m_{h_1}^{2} - m_{h_3}^{2}) }, 
  \label{eq:sin:gamma:general}
  \\
  s_\sigma^{2} &= \frac{m_{h_2}^{2} - m_{h_1}^{2}}{m_{h_2}^{2} - m_{h_3}^{2}} \; \frac{ \det M^{2} + m_{h_3}^{2} (m_{h_3}^{2} - \tr M^{2}) }
  { \det M^{2} - m_{h_2}^{2} m_{h_3}^{2} + m_{h_1}^{2} (m_{h_2}^{2} + m_{h_3}^{2} - \tr M^{2}) },
  \label{eq:sin:sigma:general}
  \end{align}
  \begin{align}
  \sin 2 \alpha&=
  \Big( 
    \pm 2 \abs{s_\gamma s_\sigma} \sqrt{1 - s_\sigma^2} \sqrt{1 - \sin^2 2 \xi} 
    \, (m_{h_3}^2 - m_{h_2}^2) 
    \label{eq:sin:2alpha:general}
    \\ 
    & + [ m_{h_3}^2 - m_{h_2}^2 s_\gamma^2 + s_\sigma^2 (1 + s_\gamma^2) (m_{h_2}^2 - m_{h_3}^2) - (1 - s_\gamma^2) m_{h_1}^2 ] \sin 2 \xi
  \Big)
  \notag \\
  & \times \Big( \left[m_{h_3}^{2} -m_{h_1}^{2} + s_\gamma^{2} (m_{h_1}^{2} - m_{h_2}^{2}) \right]^{2} 
  + (m_{h_3}^{2} - m_{h_2}^{2}) (1 - s_\gamma^{2}) s_\sigma^{2}\notag\\
   &\times\left[2 m_{h_1}^{2} (1 + s_{\gamma}^{2}) 
  - 2 (m_{h_3}^2 + s_{\gamma}^{2} m_{h_2}^2) + s_\sigma^{2} (m_{h_3}^{2} - m_{h_2}^{2}) (1 - s_\gamma^{2})\right] 
  \Big)^{-\frac{1}{2}},
  \notag
  \end{align}
where $\xi$ is the mixing angle of $M^{2}$. In the limit of $s_\gamma = s_\sigma = 0$, the angle $\xi$ goes to $\alpha$. There are two solutions for $\alpha$ given by the $\pm$ signs in \eqref{eq:sin:2alpha:general}.

These expressions for the mixing angles do not involve the unknown parameters $M_1, M_2, M_3$, which depend on the specific NMSSM. Their values in particular cases may limit the range of the physical parameters $m_{h_1,h_2,h_3}, m_{H^{\pm}}$ and $\alpha, \gamma, \sigma$ but cannot affect Eqs. (\ref{eq:sin:gamma:general}, \ref{eq:sin:sigma:general}, and \ref{eq:sin:2alpha:general}). To our knowledge, analytical expressions for the mixing angles in the general NMSSM have not been presented before.

To simplify the analysis we consider two limiting cases:
\begin{itemize}
\item $H$ decoupled: In Eq. (\ref{scalar_mass_matrix}) $m_A^2 \gg v M_1, v M_2$ or $m_{h_3} \gg m_{h_1,h_2}$ and $\sigma, \delta = \alpha - \beta +\pi/2 \rightarrow 0$,

\item Singlet decoupled: In Eq. (\ref{scalar_mass_matrix}) $M_3^2 \gg v M_1, v M_2$ or $m_{h_2} \gg m_{h_1,h_3}$ and $\sigma, \gamma \rightarrow 0$,
\end{itemize}
but we  use Eqs. (\ref{eq:sin:gamma:general}), (\ref{eq:sin:sigma:general}), and (\ref{eq:sin:2alpha:general}) to control the size of the deviations from the limiting cases when the heavier mass is lowered.


When considering the couplings of the CP-even scalars to SM particles, relevant to their production and decays, we shall not include any supersymmetric loop effect other than the one that gives rise to Eq. (\ref{delta-t}). This is motivated by the kind of spectrum outlined in the previous section, with all s-particles at their ``naturalness limit'', and provides in any event a useful well-defined reference case.  We also do not include any invisible decay of the  CP-even scalars, e.g. into a pair of neutralinos. To correct for this is straightforward with all branching ratios and signal rates that will have to be multiplied by a factor $\Gamma/(\Gamma + \Gamma_{\chi\chi})$. Finally we do not consider in this paper the two neutral CP-odd scalars, since in the general NMSSM both their masses and their composition in terms of the original fields depend upon extra parameters not related to the masses and the mixings of the CP-even states nor to the mass of the charged Higgs.

\section{$H$ decoupled}
\label{sec3}
\begin{figure}
\begin{center}
\includegraphics[width=0.48\textwidth]{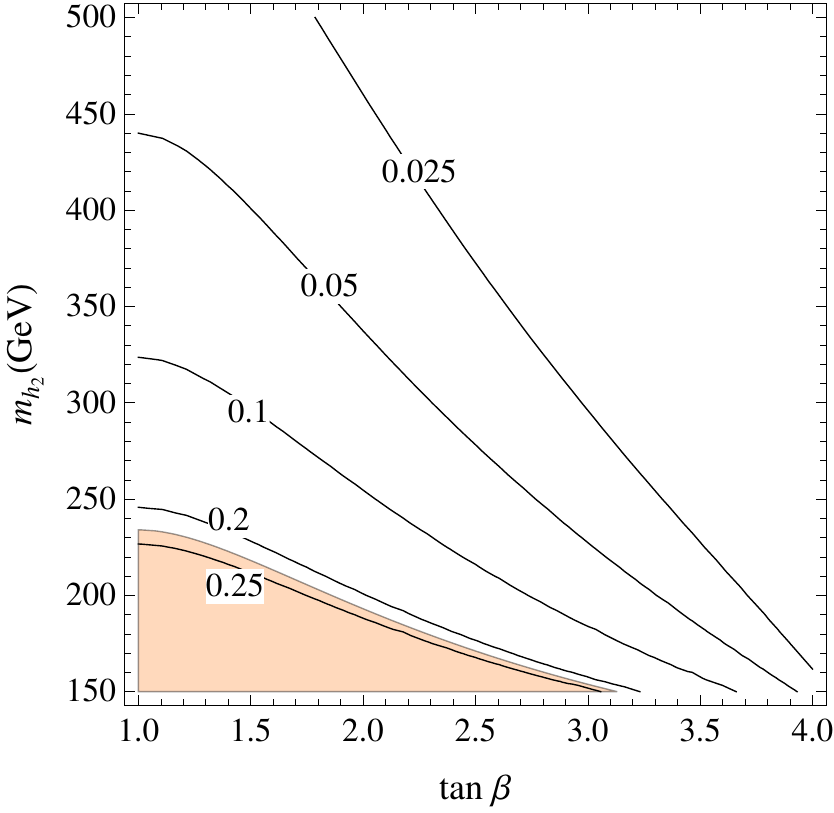}\hfill
\includegraphics[width=0.48\textwidth]{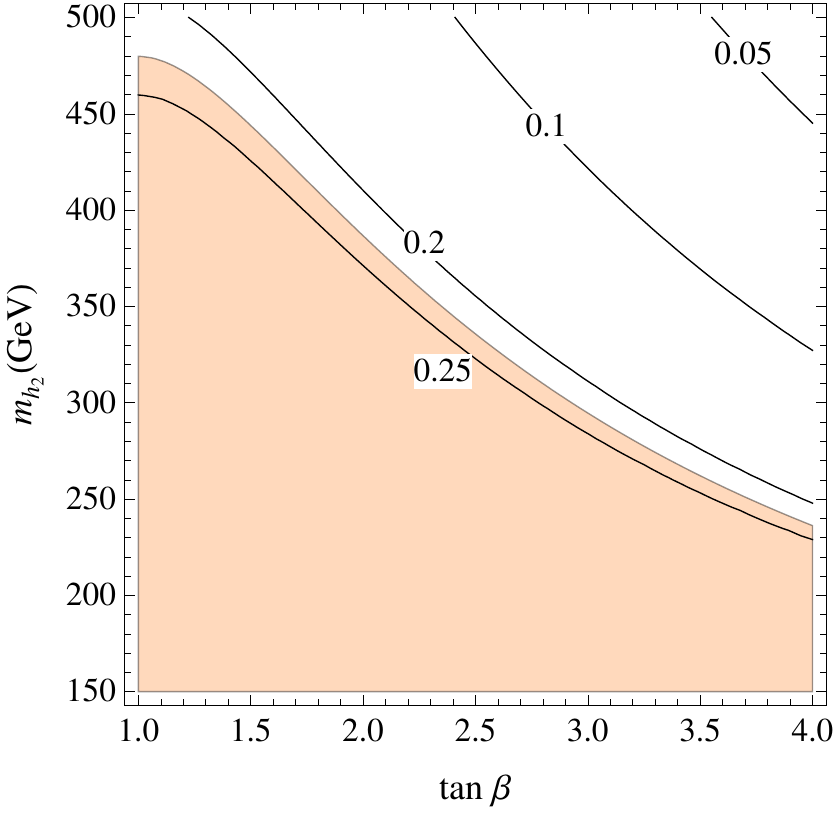}
\caption{\label{fig:mAdecoupled-1}\small $H$ decoupled. Isolines of $\sin^2\gamma$. Left: $\lambda=0.8$. Right: $\lambda=1.4$. The colored region is excluded at 95\%C.L. by the experimental data for the signal strengths of $h_1 = h_{\rm LHC}$.}
\end{center}\end{figure}

\begin{figure}
\begin{center}
\subfigure[\label{fig:mAdecoupled-Xsec8a}8 TeV, $\lambda=0.8$]{\includegraphics[width=.48\textwidth]{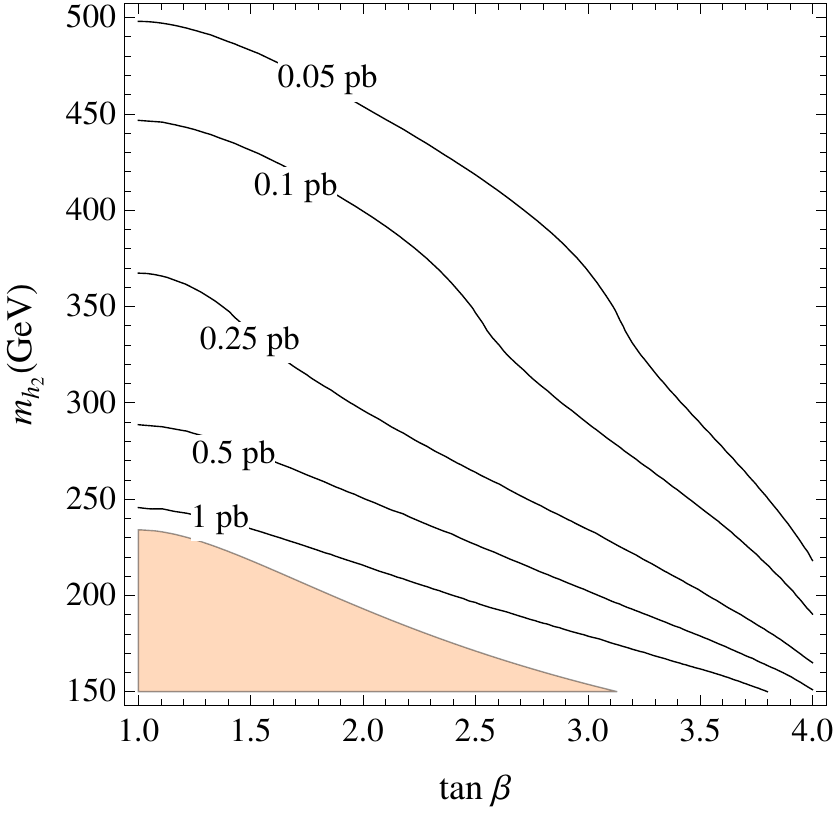}}\hfill
\subfigure[\label{fig:mAdecoupled-Xsec8b}8 TeV, $\lambda=1.4$]{\includegraphics[width=.48\textwidth]{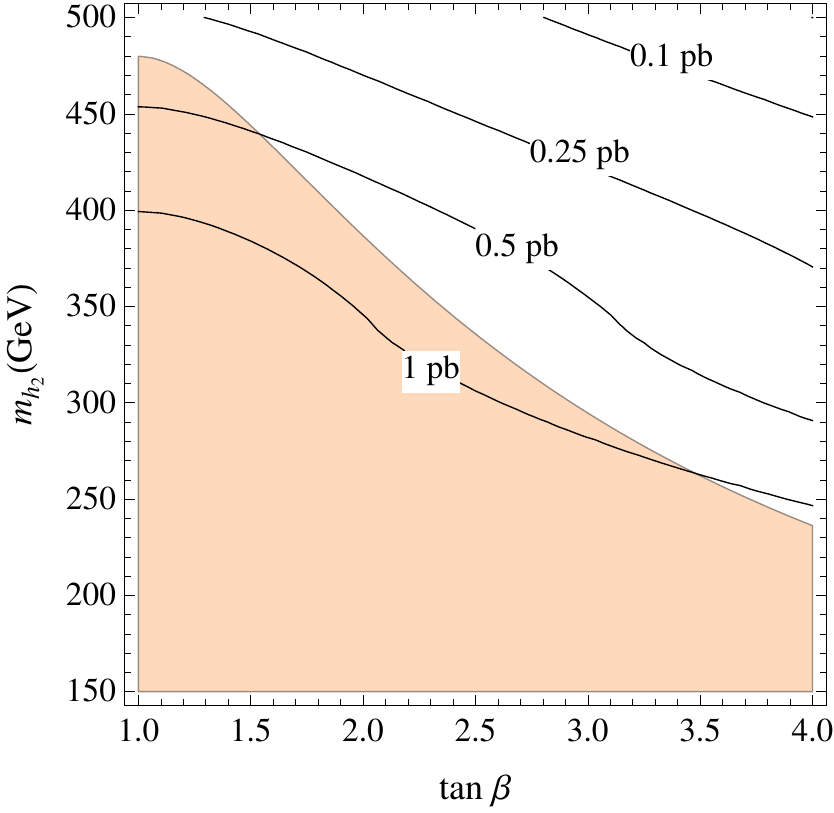}}\vspace{0.8cm}\\
\subfigure[\label{fig:mAdecoupled-Xsec14c}14 TeV, $\lambda=0.8$]{\includegraphics[width=.48\textwidth]{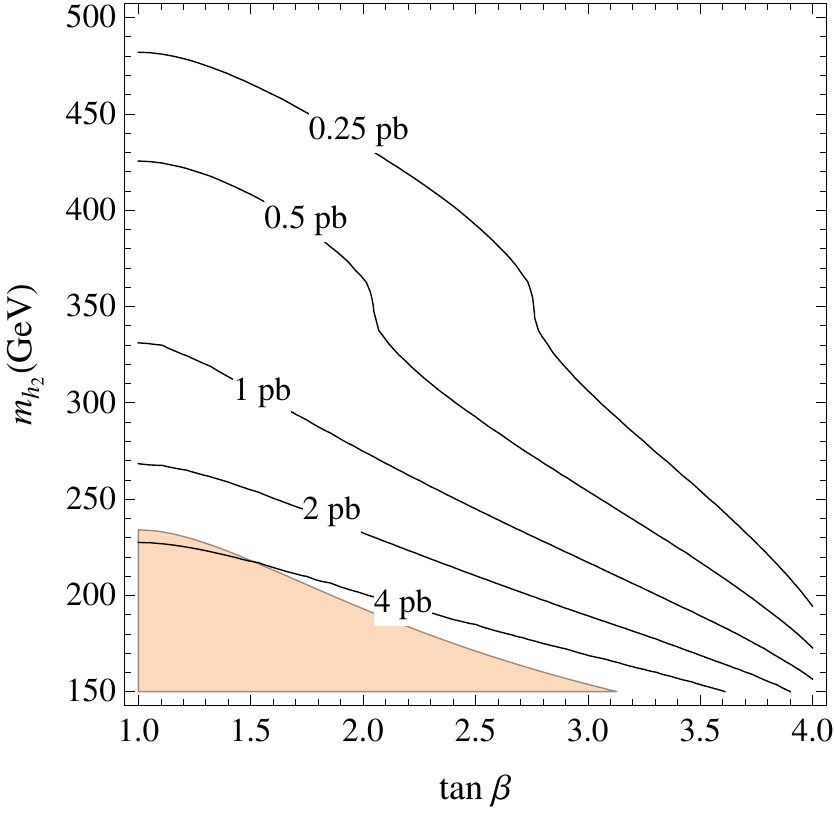}}\hfill
\subfigure[\label{fig:mAdecoupled-Xsec14d}14 TeV, $\lambda=1.4$]{\includegraphics[width=.48\textwidth]{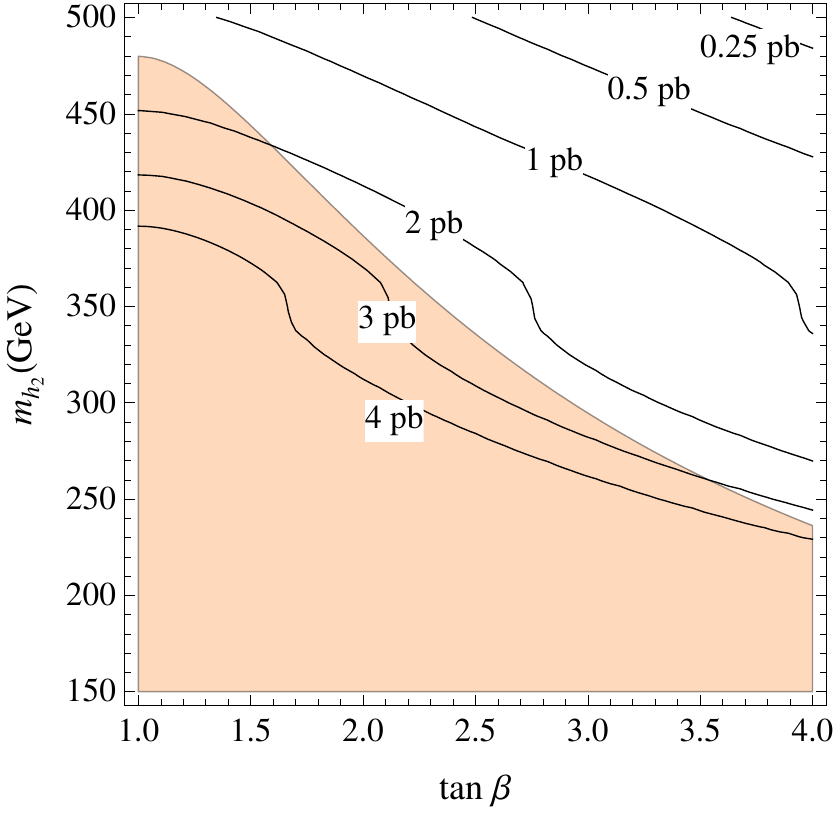}}
\caption{\label{fig:mAdecoupled-Xsec}\small $H$ decoupled. Isolines of gluon fusion cross section $\sigma(gg\to h_2)$ at LHC8 and LHC14, for the values $\lambda=0.8$ and $\lambda=1.4$. The colored region is excluded at 95$\%$C.L.}
\end{center}
\end{figure}

To study this limiting case, it is best to go in the basis $(H, h, s)$ with $H=s_\beta H_d - c_\beta H_u$ and $h=c_\beta H_d + s_\beta H_u$, and let $H\approx h_3$ decouple, so that $\sigma, \delta = \alpha - \beta +\pi/2 \rightarrow 0$. For the remaining nonvanishing angle $\gamma$ one has 
\begin{equation}
\sin^2{\gamma}= \frac{m_{hh}^2-m_{h_1}^2}{m_{h_2}^2-m_{h_1}^2},
\label{sin2gamma}
\end{equation}
where 
\begin{equation}\label{mhh}
m_{hh}^2 = m_Z^2 c_{2\beta}^2 + \lambda^2 v^2 s_{2\beta}^2 + \Delta_t^2
\end{equation}
is the first diagonal entry in the  square mass matrix of  the reduced basis $(h, s)$.

Under the conditions specified in the previous section it is straightforward to see that  the couplings of $h_1 = h_{\text{LHC}}$ and $h_2$ to fermions or to vector boson pairs, $VV = WW, ZZ$, normalized to the same couplings of the SM Higgs boson, are given by
\begin{equation}
\frac{g_{h_1ff}}{g^{\text{SM}}_{hff}} = \frac{g_{h_1VV}}{g^{\text{SM}}_{hVV}}= c_\gamma, ~~~~~
\frac{g_{h_2ff}}{g^{\text{SM}}_{hff}}= \frac{g_{h_2VV}}{g^{\text{SM}}_{hVV}}= - s_\gamma.
\end{equation}
As a consequence none of the branching ratios of $h_1$ gets modified with respect to the SM ones, whereas its production cross sections, or the various signal strengths, are reduced by a common factor $c_\gamma^2$ with respect to the SM ones with $m_{h_{\text{SM}}} = m_{h_1}$. 
The fit of all experimental data collected so far gives the bound on $s_\gamma^2$ shown in the right side of Figure~\ref{fig:FIT}.

Upon use of Eq. (\ref{sin2gamma}) the impact of this bound on the parameter space is shown in Fig.~\ref{fig:mAdecoupled-1} for $\lambda = 0.8$ and 1.4, together with the isolines of different values of $s_\gamma^2$ that might be probed by future improvements in the measurements of the $h_1$ signal strengths. Larger values of $\lambda$ already exclude a significant portion of the parameter space at least for moderate $\tan{\beta}$, as preferred by naturalness. In this section we are taking a fixed value of $\Delta_t = 73$~GeV in (\ref{delta-t}), which is obtained, e.g., for $m_{\tilde{t}_1} = 600$~GeV, $m_{\tilde{t}_2} = 750 $~GeV and mixing angle $\theta_t = 45^0$ \cite{Carena:1995bx}. As long as one stays at $\Delta_t \lesssim 85$ GeV, in a range of moderate fine tuning, and $\lambda \gtrsim 0.8$, our results do not depend significantly on $\Delta_t$.

In the same $(\tan{\beta}, m_{h_2})$ plane of Figure~\ref{fig:mAdecoupled-1} and for the same values of $\lambda$, Figure~\ref{fig:mAdecoupled-Xsec} shows the gluon-fusion production cross sections of $h_2$ at LHC for 8 or 14 TeV center-of-mass energies, where we rescaled by $c_\gamma^2$ the (next-to-next-to-leading logarithmic) NNLL ones provided in \cite{Dittmaier:2011ti}. All other $h_2$ production cross sections, relative to the gluon-fusion one, scale as in the SM with $m_{h_{\text{SM}}} = m_{h_2}$.

\begin{figure}[t!]
\begin{center}
\includegraphics[width=.48\textwidth]{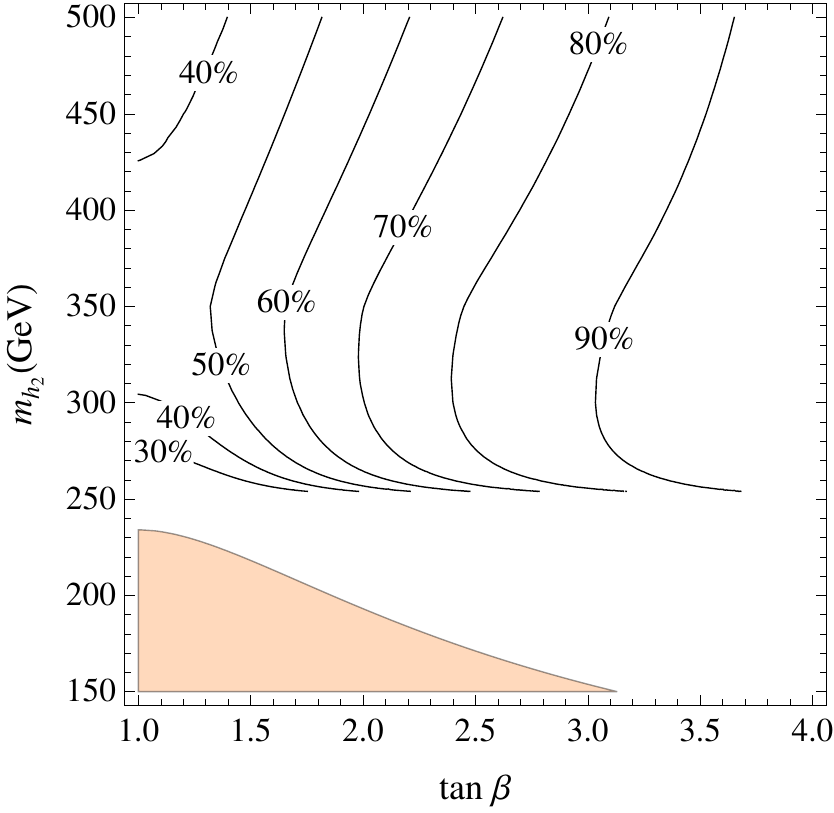}\hfill
\includegraphics[width=.48\textwidth]{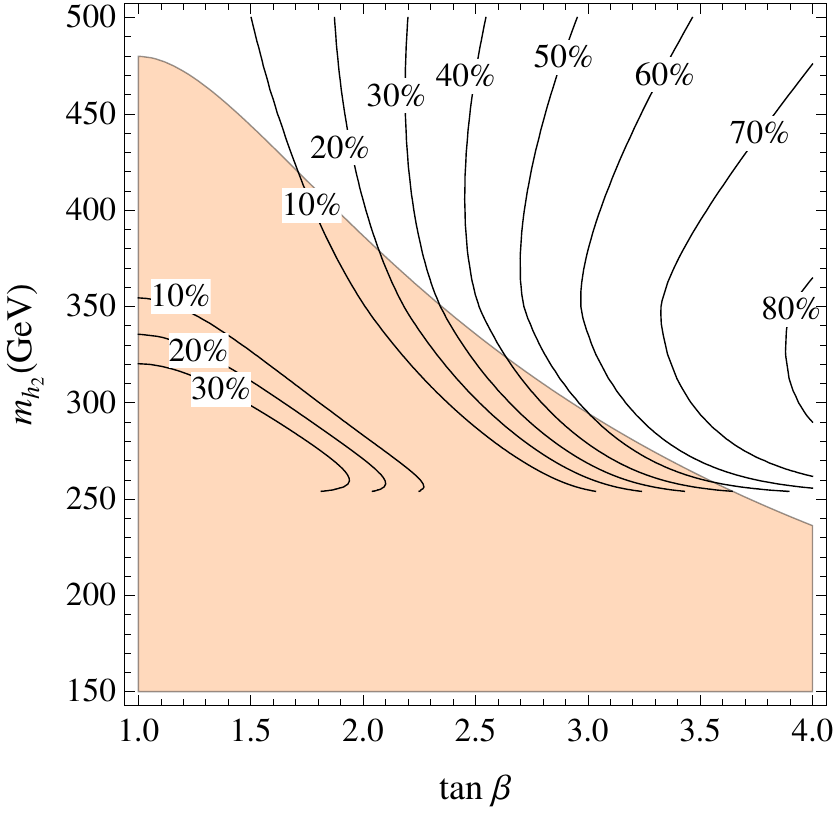}
\caption{\label{fig:mAdecoupled-hh}\small $H$ decoupled. Isolines of BR$(h_2\to h h )$. Left: $\lambda=0.8$ and $v_S=2v$. Right: $\lambda=1.4$ and $v_S=v$. The colored region is excluded at 95$\%$C.L.}
\vspace{2cm}
\includegraphics[width=.48\textwidth]{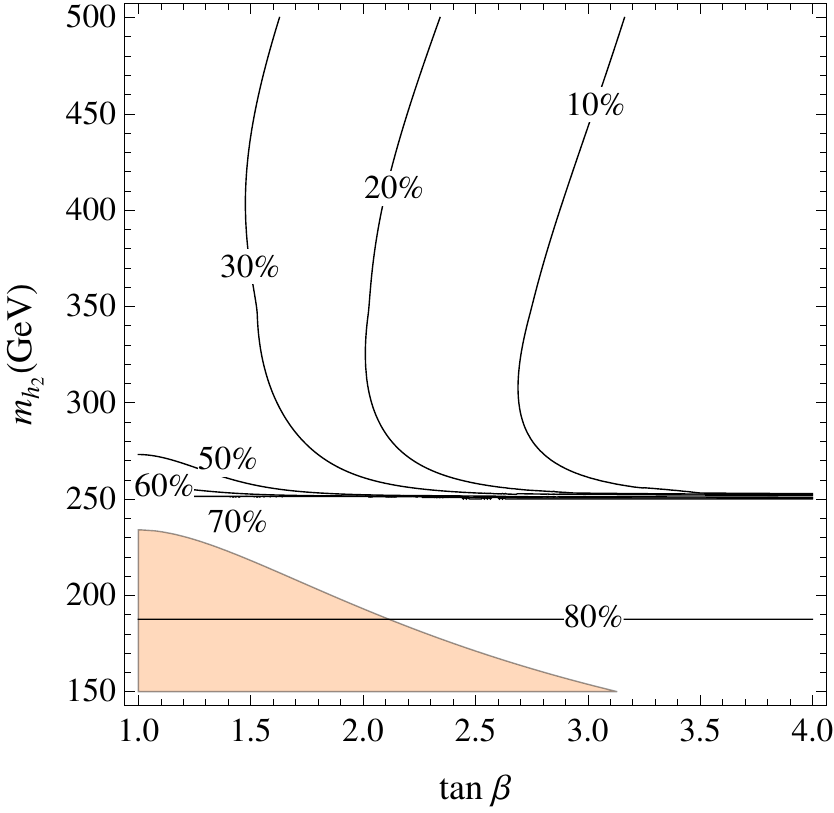}\hfill
\includegraphics[width=.48\textwidth]{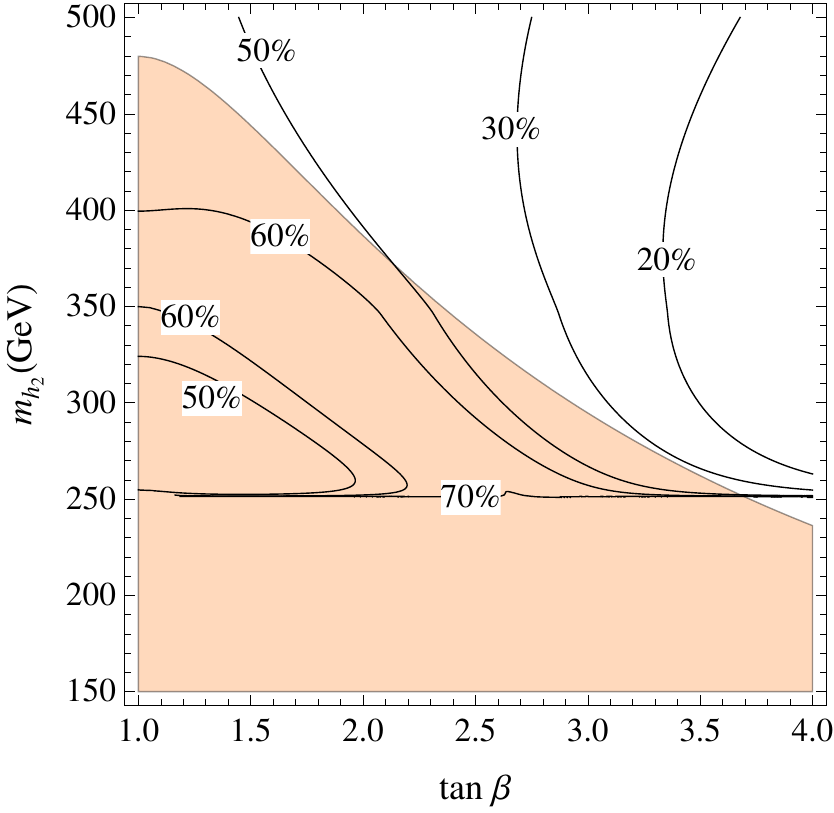}
\caption{\label{fig:mAdecoupled-WW}\small $H$ decoupled. Isolines of BR$(h_2\to W^+W^- )$. Left: $\lambda=0.8$ and $v_S=2v$. Right: $\lambda=1.4$ and $v_S=v$. The colored region is excluded at 95$\%$C.L.}
\end{center}
\end{figure}\eject
\begin{figure}[t!]
\begin{center}
\includegraphics[width=.48\textwidth]{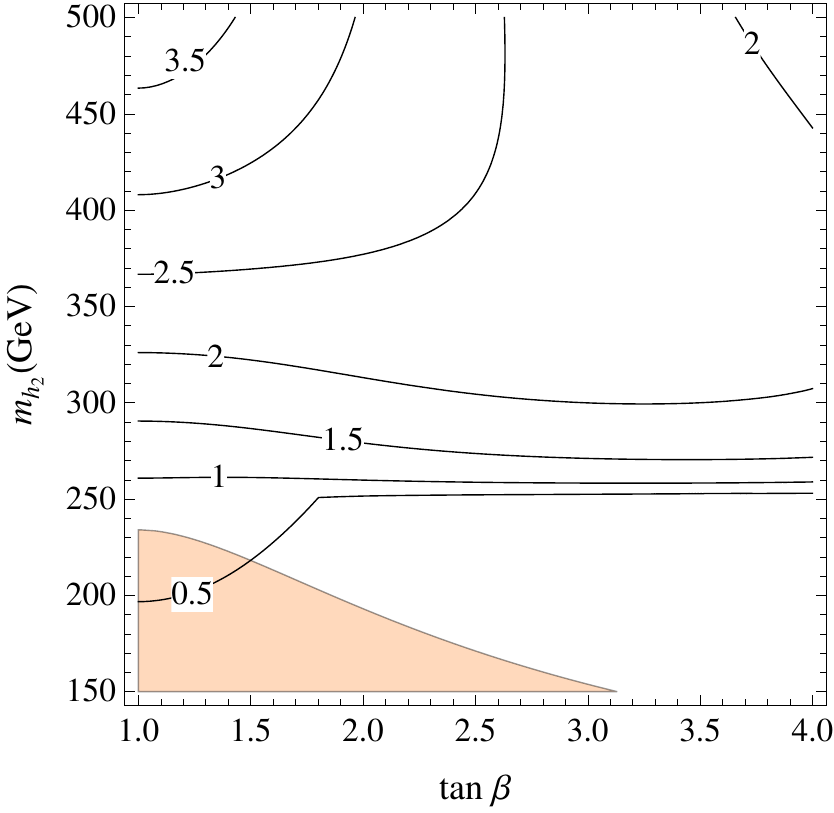}\hfill
\includegraphics[width=.48\textwidth]{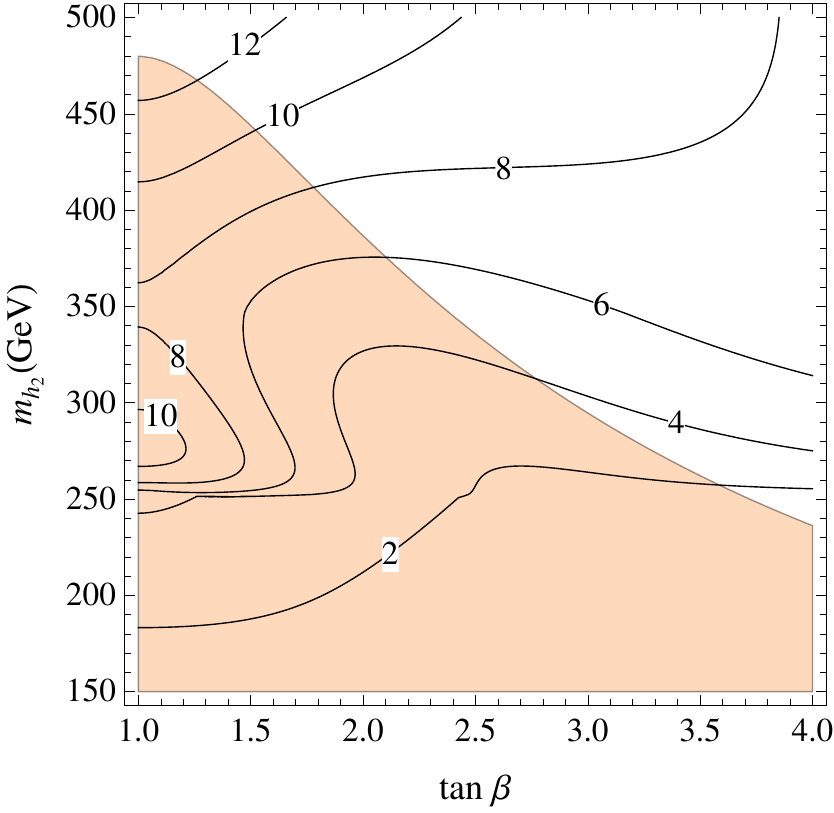}
\caption{\label{fig:mAdecoupled-width}\small $H$ decoupled. Isolines of the total width $\Gamma_{h_2}(\text{GeV})$. Left: $\lambda=0.8$ and $v_S=2v$. Right: $\lambda=1.4$ and $v_S=v$. The colored region is excluded at 95$\%$C.L.}
\end{center}
\end{figure}

\begin{figure}
\begin{center}
\includegraphics[width=.48\textwidth]{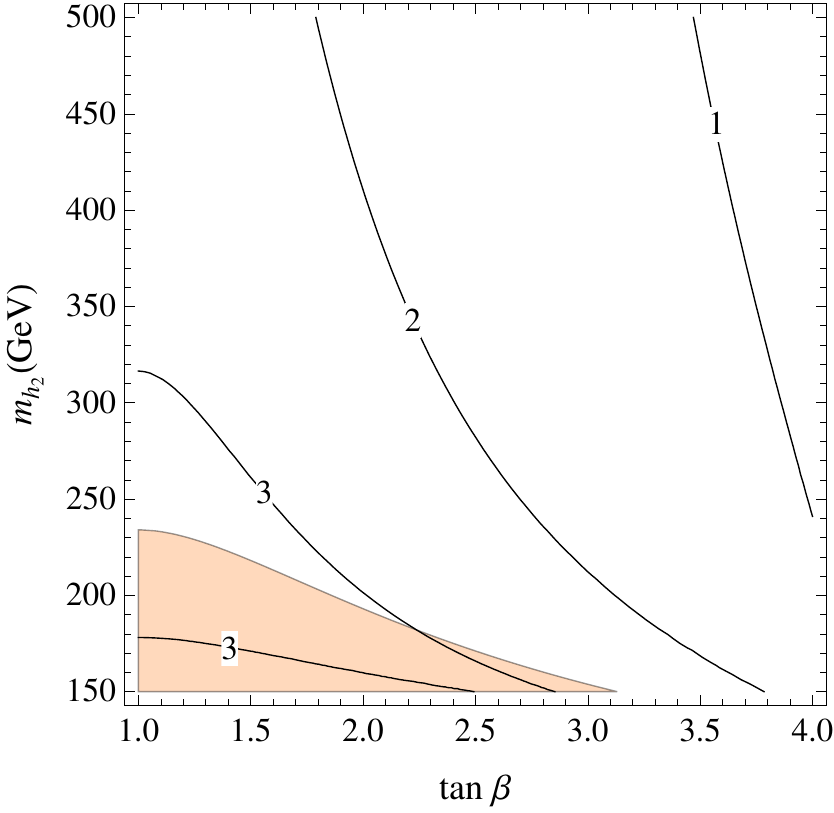}\hfill
\includegraphics[width=.48\textwidth]{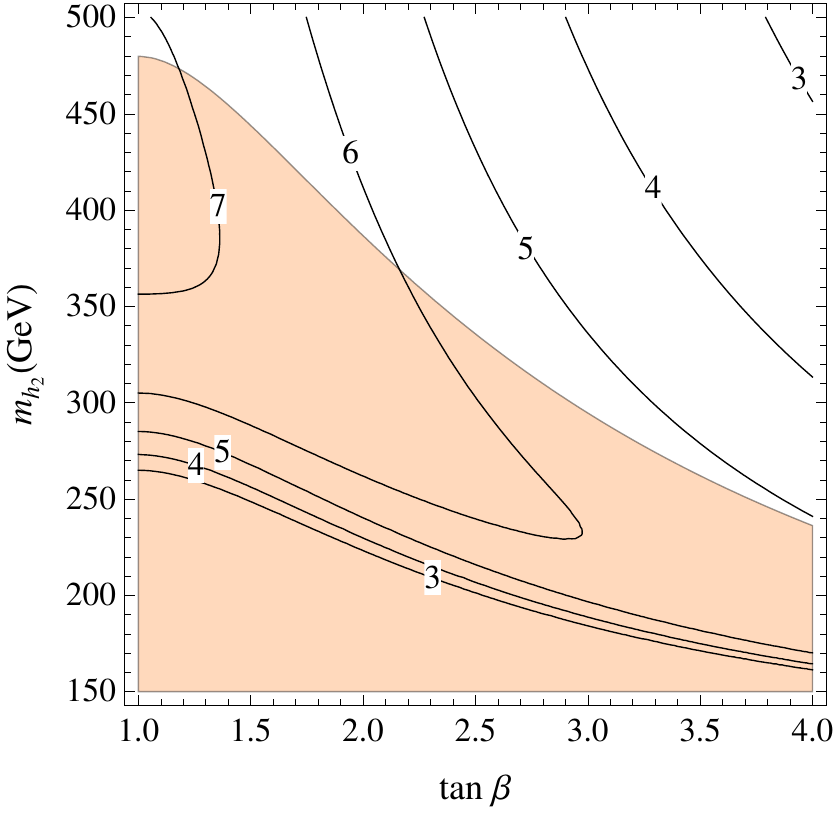}
\caption{\label{fig:mAdecoupled-cubic}\small $H$ decoupled. Isolines of $g_{hhh}/g_{hhh}^{\text{SM}}$. Left: $\lambda=0.8$ and $v_S=2v$. Right: $\lambda=1.4$ and $v_S=v$. The colored region is excluded at 95$\%$C.L.}
\vspace{2cm}
\includegraphics[width=.48\textwidth]{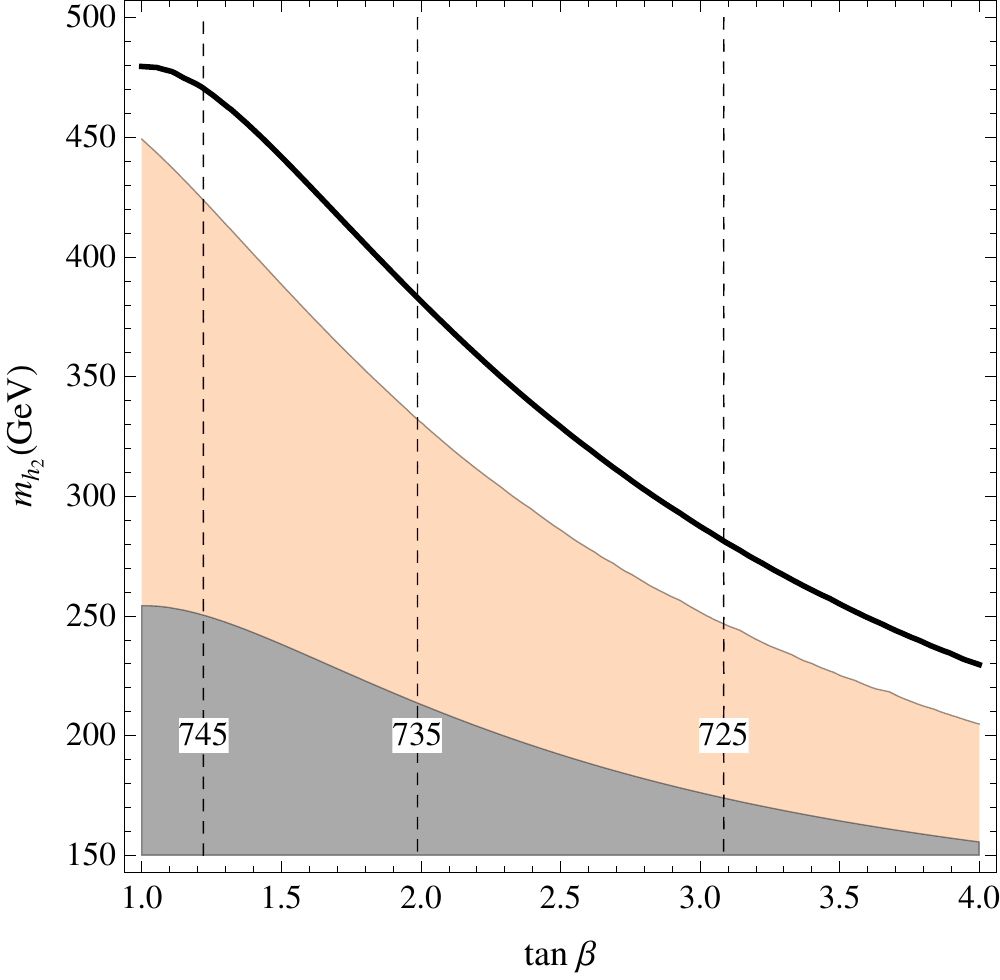}\hfill
\includegraphics[width=.48\textwidth]{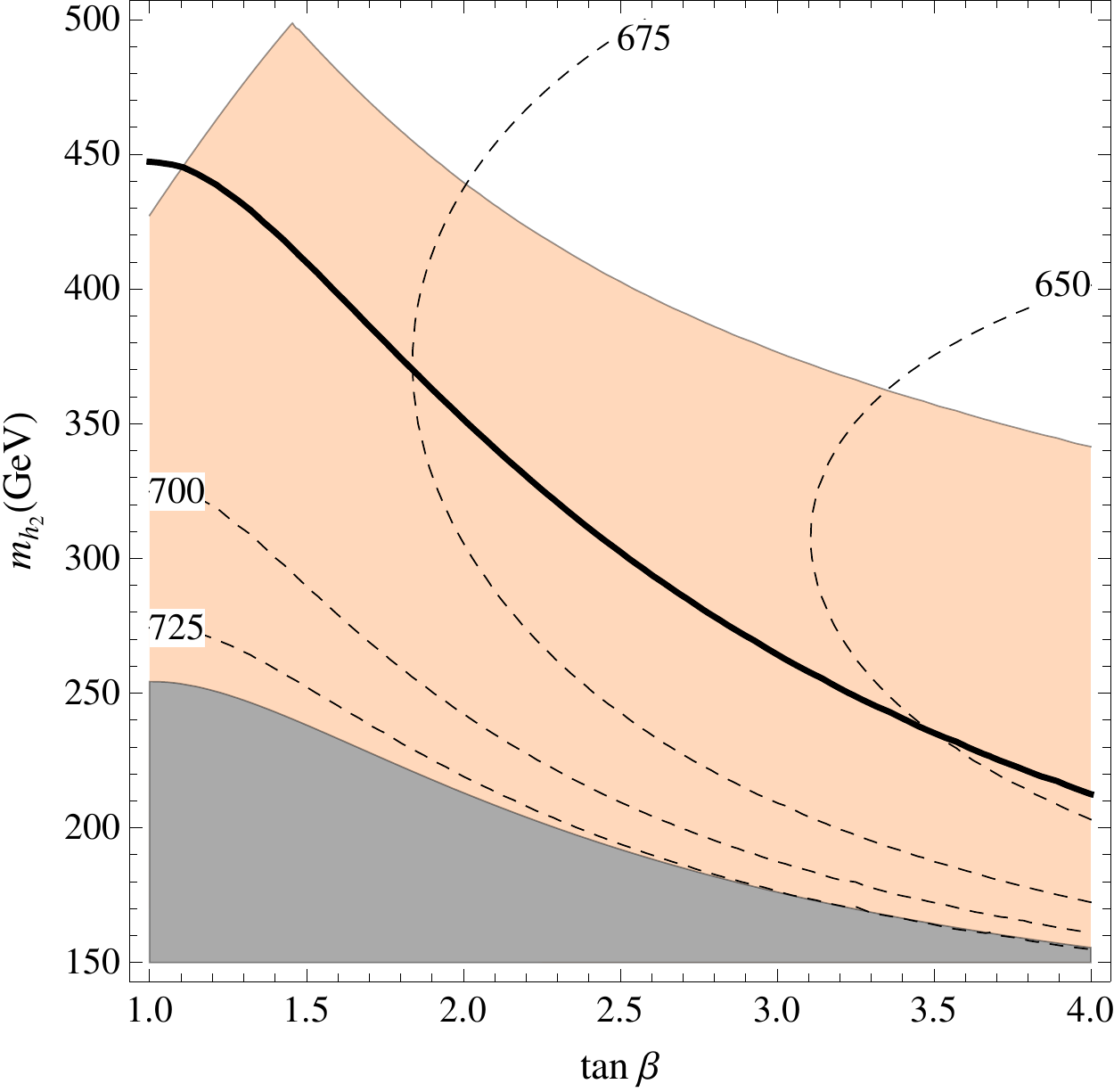}
\caption{\label{fig:mA:quasidecoupled}\small $H$ ``almost decoupled'' with $\lambda = 1.4$ and $m_{h_3} = 750$ GeV. The dashed isolines are for $m_{H^{\pm}}$. Left: $\sin^2\sigma=0$. Right: $\sin^2\sigma=0.25$. The colored region is excluded at $95\%$C.L. In the grey area there is no solution for $\delta$. The thick line shows the na\"ive exclusion limit from $s_{\gamma}^{2}$ only.}
\end{center}
\end{figure}

To determine the decay properties of $h_2$ it is crucial to know its coupling $(g_{h_2h_1^2}/2) h_2h_1^2$ to the lighter state. In the general NMSSM and in the large-$m_H$ limit considered in this section, the leading $\lambda^2$-term contribution to this coupling, as well as the one to the cubic $h_1$ coupling $(g_{h_1^3}/6) h_1^3$, are given by
\begin{align}
g_{h_2h_1^2}&=\frac{\lambda ^2 v}{8 \sqrt{2}} \left(4 \frac{v_S}{v} \cos\gamma +12 \frac{v_S}{v} \cos3 \gamma-7  \sin\gamma +12  \cos4 \beta \cos^2\gamma \sin\gamma +9  \sin3 \gamma \right)\notag\\
&- \frac{3}{\sqrt{2}v}\Delta_t^2 \cos^2\gamma\sin\gamma,\\
\frac{g_{h_1^3}}{g^{\text{SM}}_{h_1^3}}&=\frac{\lambda ^2 v^2}{8 m_{h_1}^2} \cos\gamma \left(10 - 4  \cos4 \beta \cos^2\gamma - 6 \cos2 \gamma + 8 \frac{v_S}{v} \sin2 \gamma\right) + \dfrac{\Delta_t^2}{m_{h_1}^2} \cos^3\gamma,
\end{align}
where $v_S$ is the VEV of the singlet
. Figures~\ref{fig:mAdecoupled-width} and \ref{fig:mAdecoupled-hh} show the total width of $h_2$ and its branching ratio into a pair of light states for some choices of $v_S$. 
The other most significant decay mode of $h_2$ is into a $W$ pair, with a branching ratio given in Figure~\ref{fig:mAdecoupled-WW}.  Figure~\ref{fig:mAdecoupled-cubic} shows the triple $h_1$ coupling normalized to the SM one.

These results depend on the value taken by $v_S$, in particular we note that the Higgs fit still allows the triple Higgs coupling to get a relative enhancement of a factor of a few (with a negative or positive sign) with respect to the Standard Model one, thus yielding potentially large effects in Higgs pair production cross sections\cite{Baglio:2012np}.


In the general case, when $H$ is not completely decoupled, the three angles $\delta$, $\gamma$ and $\sigma$ can all be different from zero, and the three masses $m_{h_2}$, $m_{h_3}$ and $m_{H^\pm}$ are all virtually independent.
In Figure~\ref{fig:mA:quasidecoupled} we show the excluded regions in the plane $(\tan\beta,m_{h_2})$ for $m_{h_3} = 750$ GeV and $\lambda = 1.4$, setting $s^2_{\sigma}$ to two different values in order to fix $m_{H^\pm}$. When $s^2_{\sigma} = 0$ one recovers the previous $H$ decoupled case in the limit $m_{h_3}\to\infty$. With respect to this case, both $\gamma$ and $\delta$ are free parameters in the fit to the couplings of $h_{\rm LHC}$, and as a consequence the bounds are milder than what is expected from using only $\gamma$. If $s^2_{\sigma} \neq 0$, $h_2$ and $h_3$ are not decoupled, and their masses can not be split very consistently with all the other constraints. This is reflected in a broader excluded region for low $m_{h_2}$ in the right side of Figure~\ref{fig:mA:quasidecoupled}, where we take $s^2_{\sigma} = 0.25$.


\section{Singlet decoupled}
\label{sec4}
\begin{figure}[t]
\begin{center}
\includegraphics[width=.48\textwidth]{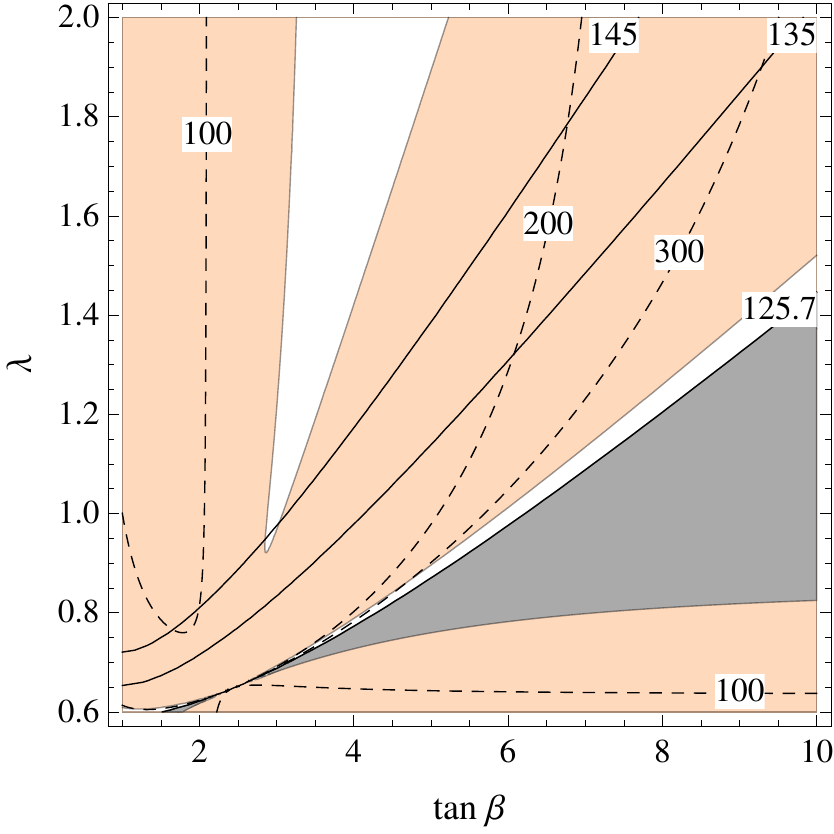}\hfill
\includegraphics[width=.48\textwidth]{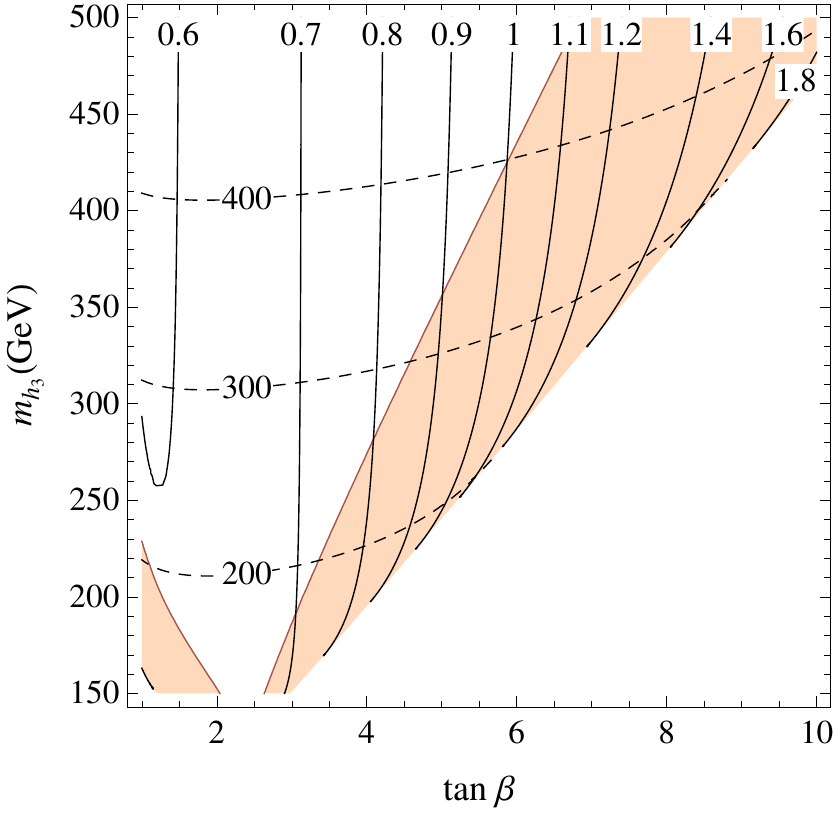}
\caption{\label{fig:mSdecoupled-1}\small Singlet decoupled. Left: Isolines of $m_{hh}$ \eqref{mhh} (solid), the grey region is unphysical due to $m_{H^{\pm}}^{\,2} < 0$. Right: Isolines of $\lambda$ in the region with $m_{hh}\simeq 126$ GeV. The dashed isolines are for $m_{H^{\pm}}$. The colored regions are excluded at 95$\%$C.L.}
\end{center}
\end{figure}
Let us consider now the limit in (\ref{scalar_mass_matrix}) $M_3^2 \gg v M_1, v M_2$, which corresponds to $m_2 \gg m_{1,3}$ and $\sigma, \gamma \rightarrow 0$. 
In this case the three relations that have led to Eqs. (\ref{eq:sin:gamma:general}), (\ref{eq:sin:sigma:general}), (\ref{eq:sin:2alpha:general}) become
\begin{align}
\sin 2\alpha &= \sin 2\beta \; \frac{2\lambda^2 v^2-m_Z^2-m_A^2|_{m_{h_1}}}{m_A^2|_{m_{h_1}} +m_Z^2 +\delta_t^2 -2m_{h_1}^2},\\
m_{h_3}^2&= m_A^2|_{m_{h_1}}+m_Z^2 +\delta_t^2 -m_{h_1}^2,
\end{align}
where
\begin{equation}\label{mA_mh}
m_A^2\big|_{m_{h_1}}=\frac{\lambda^2v^2(\lambda^2v^2-m_Z^2)\sin^2 2\beta-m_{h_1}^2(m_{h_1}^2-m_Z^2-\delta_t^2)-m_Z^2\delta_t^2 \cos^2\beta}{m_{hh}^2-m_{h_1}^2}.
\end{equation}
Note the difference with respect to the single relation (\ref{sin2gamma}) of the previous section. Identifying as before $h_1$ with the resonance found at the LHC, this determines  the mass of $h_3$ (and of  $H^{\pm}$) for any given value of 
$\lambda$ and $\tan{\beta}$. Still taking $\Delta_t= 73$ GeV, this  allows to show plots similar to the ones of the previous section without having to fix the value of $\lambda$. 

Defining as before $\delta = \alpha - \beta +\pi/2$, the couplings of  $h_3$ become
\begin{equation}
\frac{g_{h_3tt}}{g^{\text{SM}}_{htt}}=\sin\delta-\frac{\cos\delta}{\tan\beta},~~\frac{g_{h_3bb}}{g^{\text{SM}}_{hbb}}=\sin\delta+\tan\beta\cos\delta,~~\frac{g_{h_3VV}}{g^{\text{SM}}_{hVV}}= \sin\delta.
\label{h3couplings}
\end{equation}
The allowed regions in the plane $(\tan{\beta}, \lambda)$ shown in the left side of Figure~\ref{fig:mSdecoupled-1} are determined by a 2-parameter fit of $\tan \beta$, $\sin \delta$. This fit results in an allowed region which is virtually the same as the one with $\gamma = 0$ in the left side of Figure~\ref{fig:FIT}. When inverting $\lambda$ as a function of $\tan\beta, m_{h_3}$, there are two solutions. In the right side of Figure~\ref{fig:mSdecoupled-1}, we show only the one which corresponds to the narrow allowed region with $m_{hh}$ close to 126 GeV. Note that $\lambda$ is restricted to relatively small values. As a consequence the analysis becomes more sensitive to values of $\Delta_t$ at or above 80 GeV.

\begin{figure}[t]
\begin{center}
\includegraphics[width=.48\textwidth]{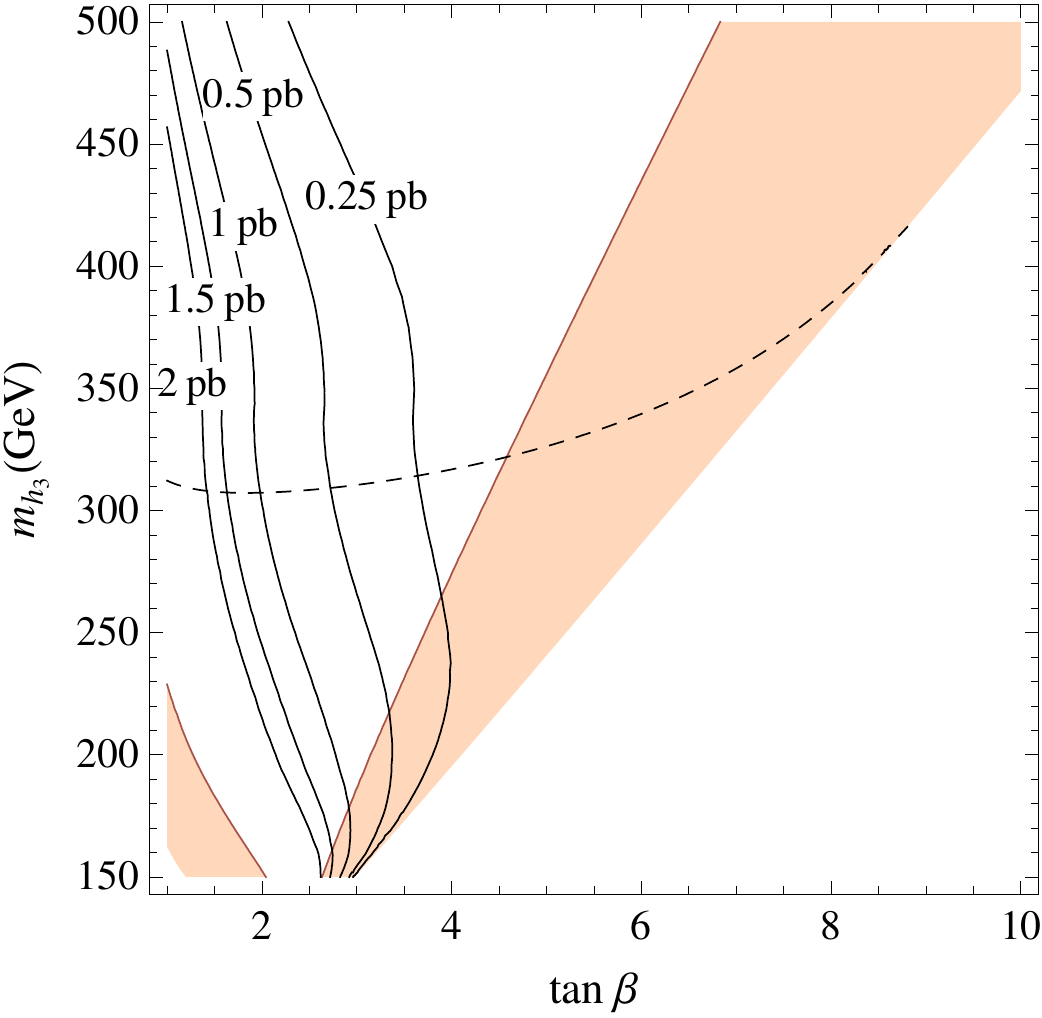}\hfill
\includegraphics[width=.48\textwidth]{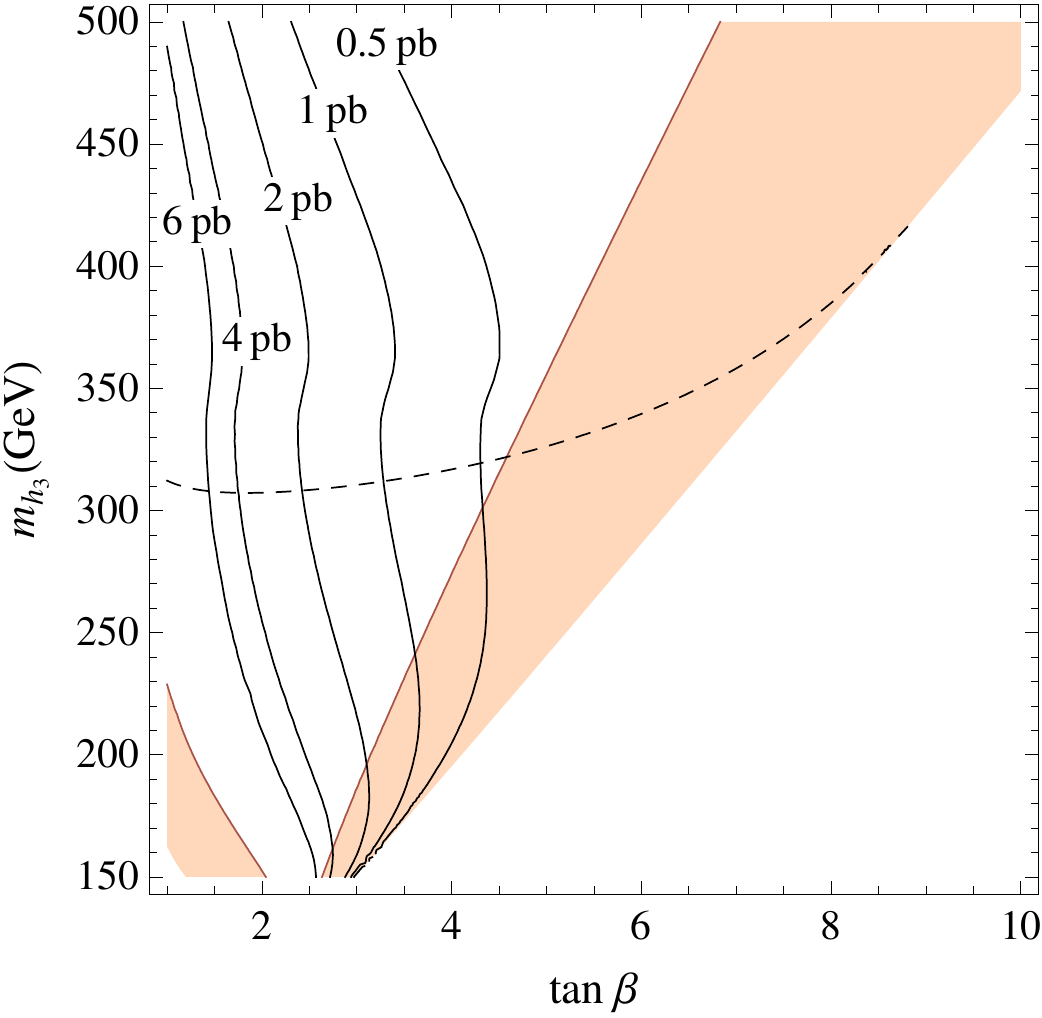}
\caption{\label{fig:mSdecoupled-Xsec}\small Singlet decoupled. Isolines of gluon fusion production cross section $\sigma(gg\to h_3)$. The colored regions are excluded at 95$\%$C.L., and the dashed line shows $m_{H^\pm}=300$ GeV. Left: LHC8. Right: LHC14.}
\end{center}
\end{figure}

The other allowed region in the left side of Figure \ref{fig:mSdecoupled-1}, when translated to the ($\tan\beta, m_{h_3}$) plane, corresponds to the other solution for $\lambda$, and is not displayed in the right side of Figure \ref{fig:mSdecoupled-1}. It always implies a charged Higgs mass $m_{H^{\pm}}$ below 150 GeV, which is disfavored by indirect constraints \cite{Misiak:2006zs}. Note that this region, corresponding to the allowed region with large $\delta$ in Figure~\ref{fig:FIT}, is mainly allowed because of the large error in the measurement of the $b\bar b$ coupling of $h_{\text{LHC}}$. Reducing this error down to about 30\% around $g_{h_1bb}/g_{hbb}^{\rm SM}\simeq 1$ would exclude the region.
As already said, we do not consider here the region with $m_{hh} < 126$ GeV, which requires a CP-even state lighter than $h_{\rm LHC}$.

The couplings (\ref{h3couplings}) allow to compute the gluon-fusion production cross section of $h_3$ by means of \cite{Cheung:2013bn}
\begin{equation}
\sigma(gg\rightarrow h_3) = \sigma^{\text{SM}}(gg\rightarrow H(m_{h_3}))
\Big|\mathcal{A}_t \frac{g_{h_3tt}}{g^{\text{SM}}_{htt}} + \mathcal{A}_b \frac{g_{h_3bb}}{g^{\text{SM}}_{hbb}}\Big|^2,
\end{equation}
where
\begin{equation}
\mathcal{A}_{t,b} =  \frac{F_{\frac{1}{2}}(\tau_{t,b})}{F_{\frac{1}{2}}(\tau_t) + F_{\frac{1}{2}}(\tau_b)},   ~~~~~~\tau_i = 4 \frac{m_i^2}{m_{h_3}^2},
\end{equation}
and $F_{\frac{1}{2}}(\tau)$ is a one-loop function that can be found e.g. in \cite{Azatov:2012qz,Carmi:2012in}. This cross section is shown in Figure~\ref{fig:mSdecoupled-Xsec}, where we used the values of $\sigma^{\text{SM}}$ at NNLL precision provided in \cite{Dittmaier:2011ti}, and the running masses $m_{t,b}$ at NLO precision. We checked the validity of this choice by performing the same computation both with the use of masses at LO precision and K-factors \cite{Anastasiou:2009kn}, and with the program \texttt{HIGLU} \cite{Spira:1995rr,Spira:1995mt}, finding in both cases an excellent agreement.

\begin{figure}[t!]
\begin{center}
\includegraphics[width=.48\textwidth]{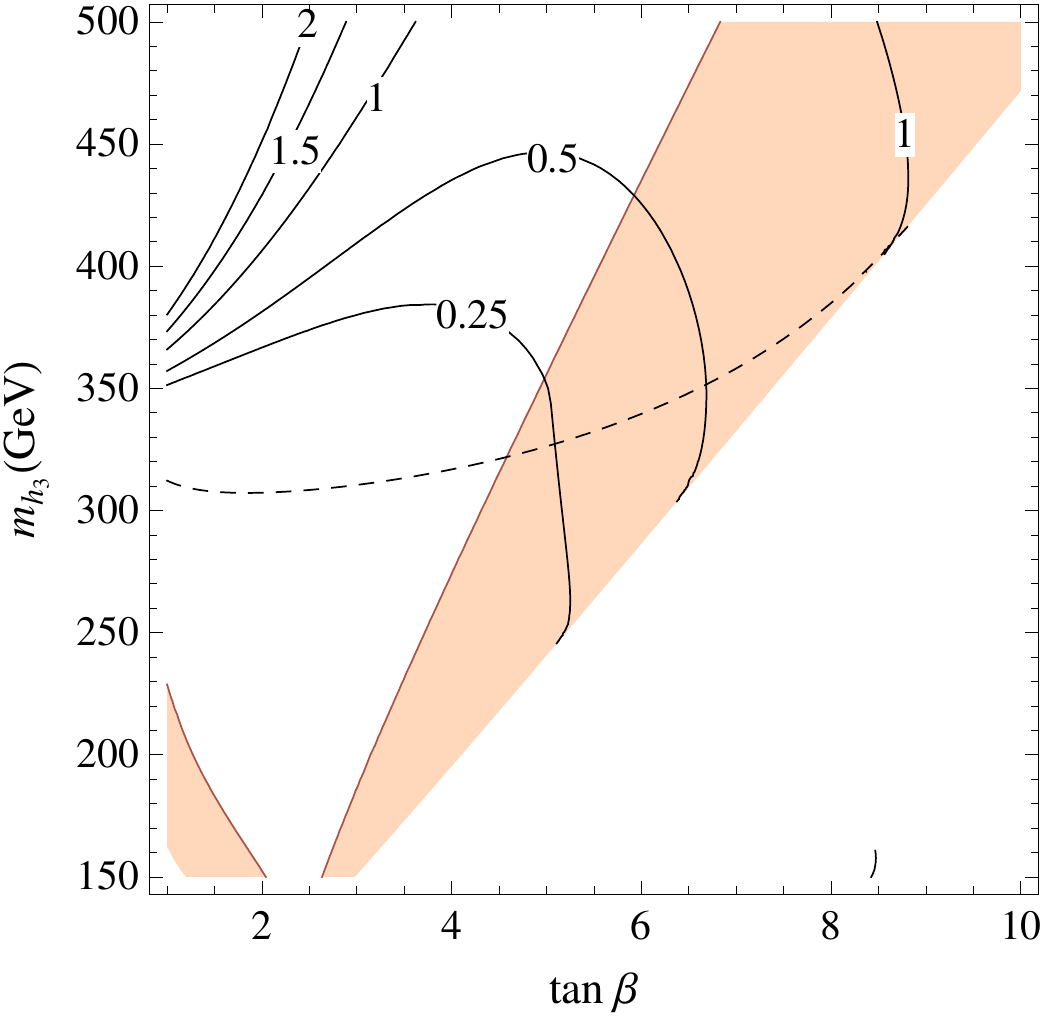}\hfill
\includegraphics[width=.48\textwidth]{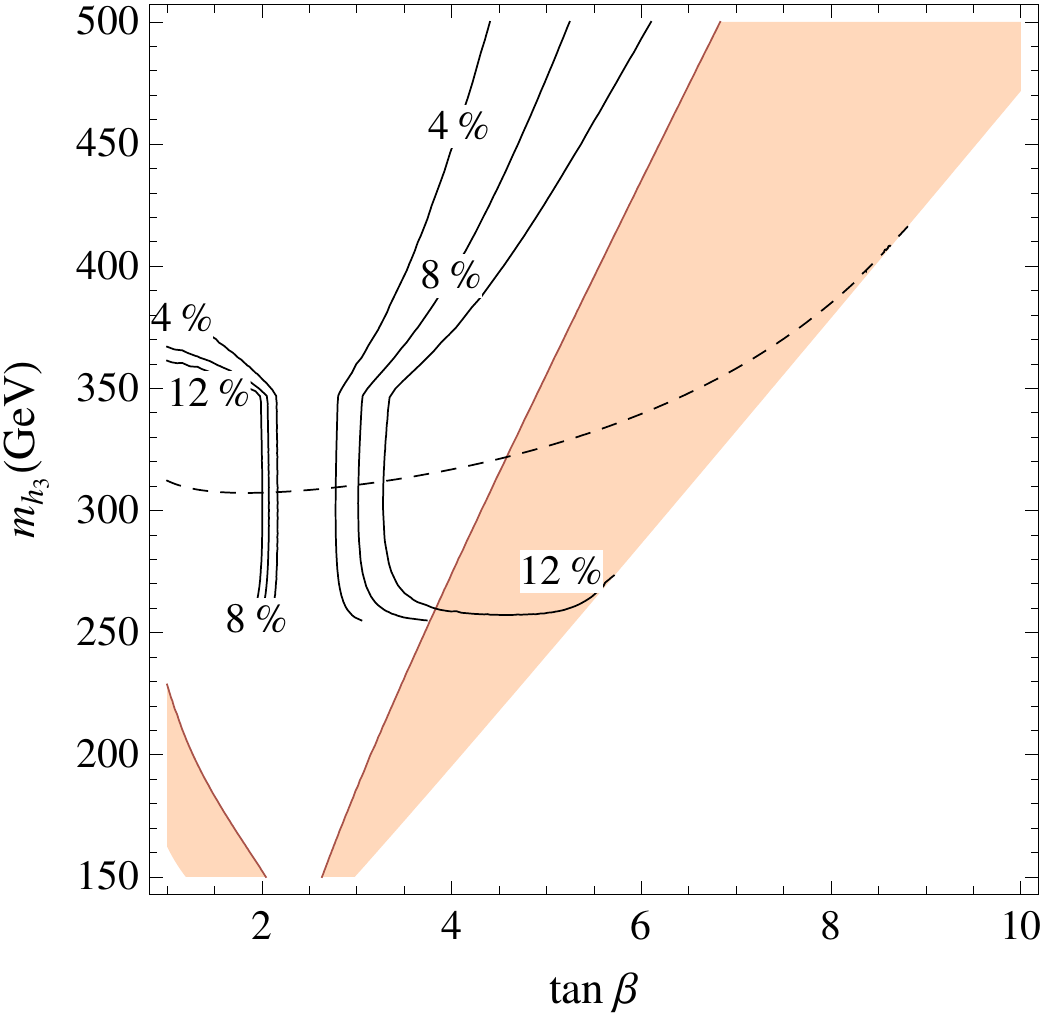}
\caption{\label{fig:mSdecoupled-BRs}\small Singlet decoupled. Left: isolines of the total width $\Gamma_{h_3}$(GeV). Right: isolines of BR$(h_3\!\to \!h h)$. The colored regions are excluded at 95$\%$C.L., and the dashed line shows $m_{H^\pm}=300$ GeV.}
\vspace{0.5cm}
\includegraphics[width=.48\textwidth]{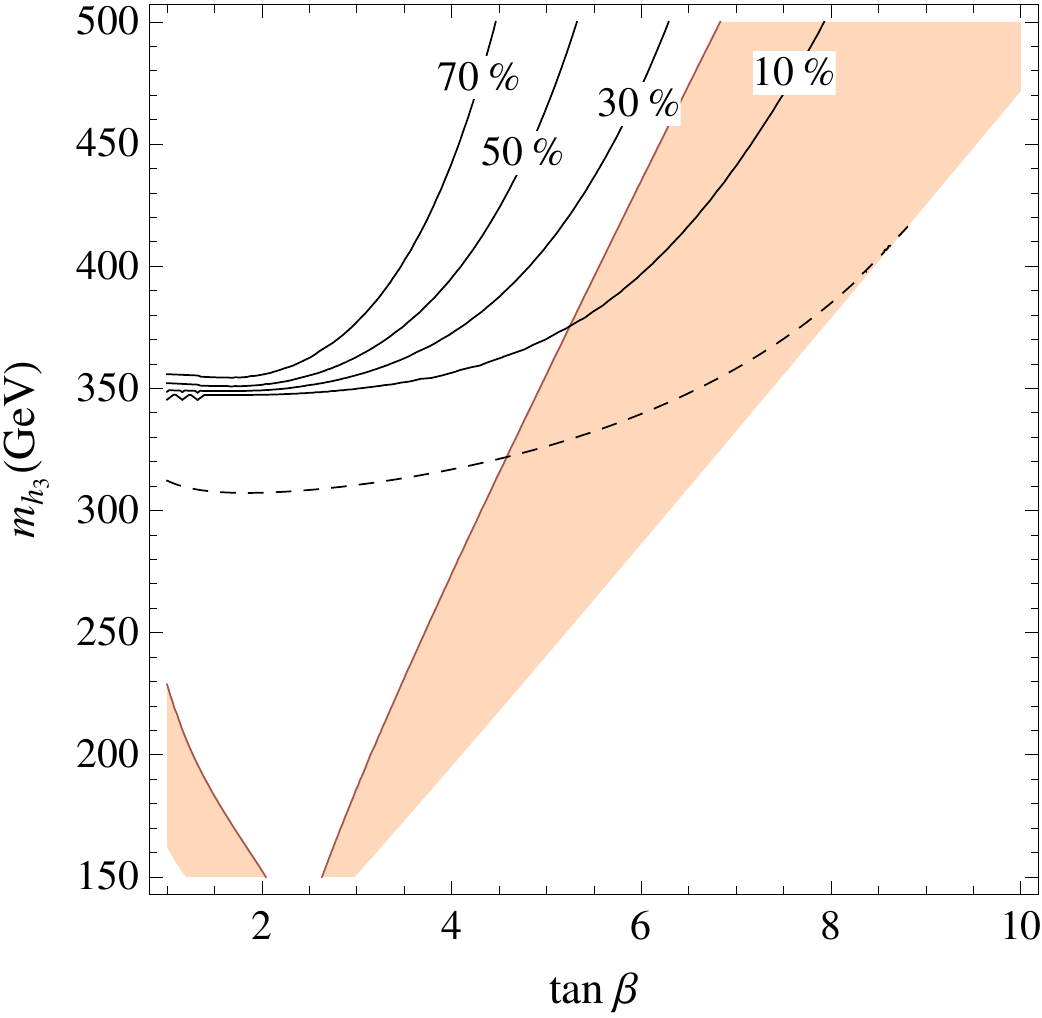}\hfill
\includegraphics[width=.48\textwidth]{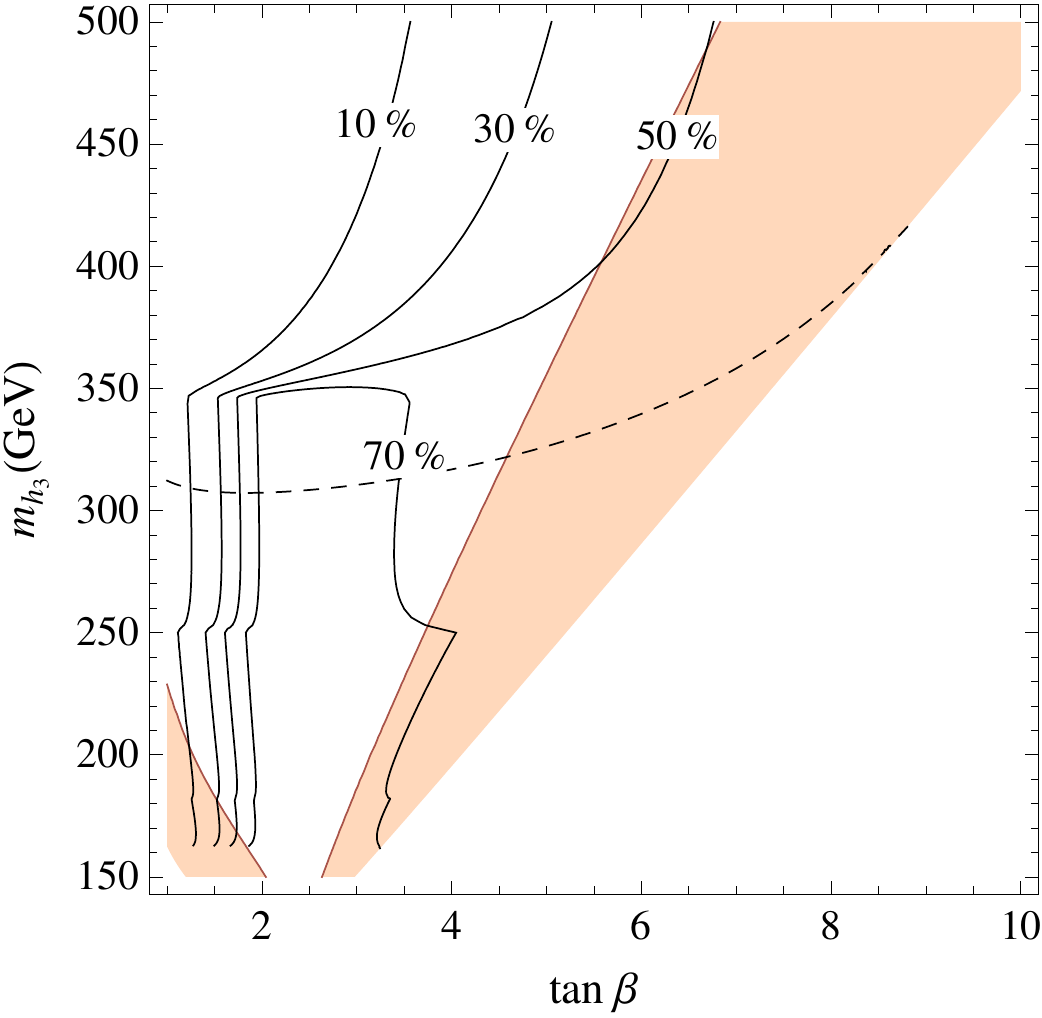}
\caption{\label{fig:mSdecoupled-BRf}\small Singlet decoupled. Left: Isolines of BR$(h_3\to t\bar t)$. Right: Isolines of BR$(h_3\to b \bar{b})$. The colored regions are excluded at 95$\%$C.L., and the dashed line shows $m_{H^\pm}=300$ GeV.}
\end{center}
\end{figure}

\noindent The coupling of $h_3$ to the lighter state $\displaystyle\frac{g_{h_3 h_1^2}}{2} h_3 h_1^2$ and the triple Higgs coupling $\displaystyle\frac{g_{h_1^3}}{6}h_1^3$ are given by
\begin{align}
g_{h_3h_1^2}&=\frac{1}{2 \sqrt{2} v} \left[(m_Z^2 + v^2 \lambda^2) \sin \delta + 3 (m_Z^2 - \lambda^2 v^2) \sin(4 \beta + 3 \delta)\right] \nonumber \\
 &- \frac{3 \Delta_t^2}{\sqrt{2} v} \frac{\cos(\beta + \delta) \sin^2(\beta+\delta)}{\sin^3\beta},\\
\frac{g_{h_1^3}}{g^{\text{SM}}_{h_1^3}} &= \frac{(m_Z^2 + v^2 \lambda^2) \cos \delta + (m_Z^2 - v^2 \lambda^2) \cos(4 \beta + 3 \delta)}{2 m_{h_1}^2} + \frac{\Delta_t^2}{m_{h_1}^2} \frac{\sin^3(\beta+\delta)}{\sin^3\beta}.
\end{align}

Figures~\ref{fig:mSdecoupled-BRs} and \ref{fig:mSdecoupled-BRf} show the most relevant widths of $h_3$.

\section{The MSSM for comparison}
\label{sec5}

\begin{figure}[t]
\begin{center}
\includegraphics[width=.48\textwidth]{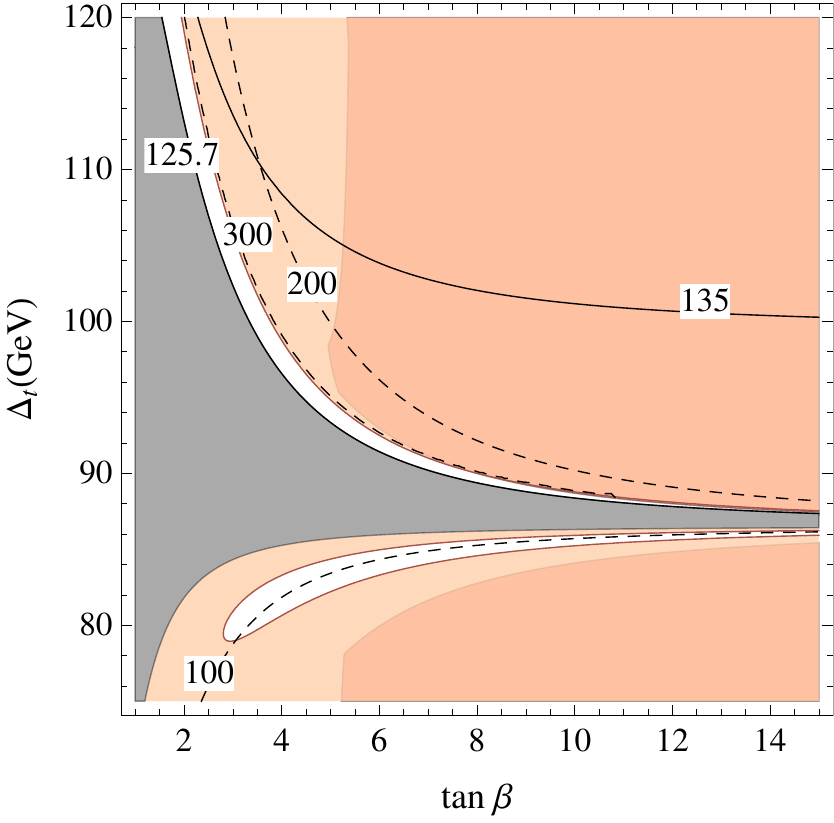}\hfill
\includegraphics[width=.48\textwidth]{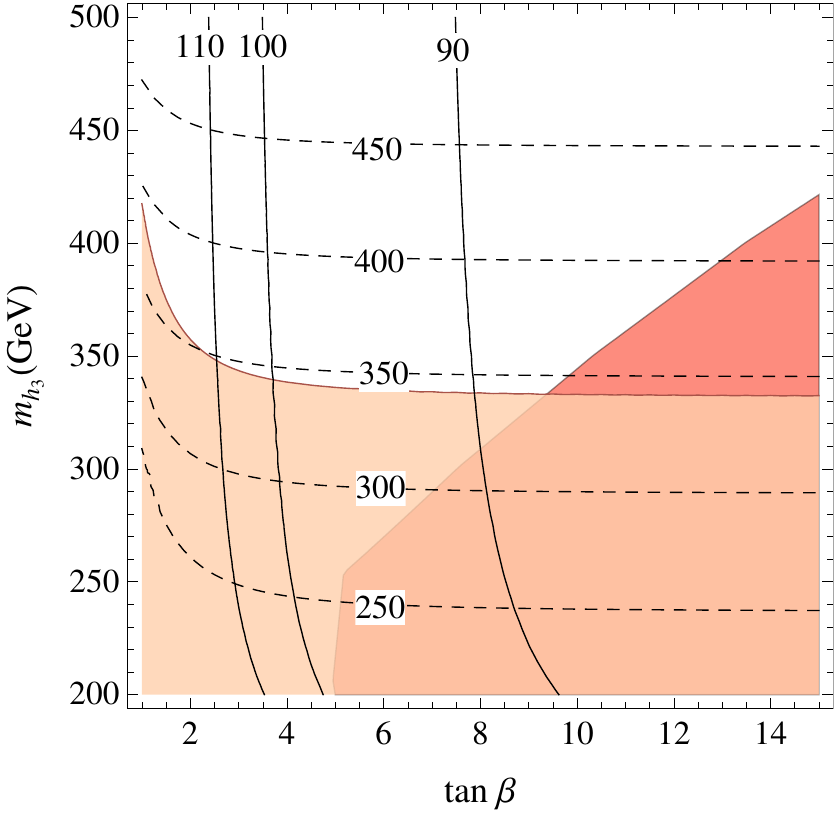}
\caption{\label{fig:MSSM-1}\small MSSM. Left: Isolines of $m_{hh}$ (solid) and $m_{H^\pm}$ (dashed), the gray region is unphysical because of $m^2_A<0$. Right: Isolines of $\Delta_t $(GeV) (solid) and $m_{H^\pm}$ (dashed). Light colored regions are excluded at 95$\%$C.L. by the Higgs fit, the red region is excluded by CMS direct searches for $A,H\to \tau^+\tau^-$.}
\end{center}
\end{figure}

It is instructive to compare the results of the previous section with the much-studied case of the MSSM, using the same language as much as possible, since the MSSM is the $\lambda=0$ limit of the NMSSM in the singlet-decoupled case. A recent paper \cite{Djouadi:2013vqa} analyzed the Higgs system of the MSSM in a way similar to ours and gave comments about the heavy Higgs searches in different channels (see also \cite{Maiani:2012ij,Maiani:2012qf}).

\begin{figure}
\begin{center}
\includegraphics[width=.48\textwidth]{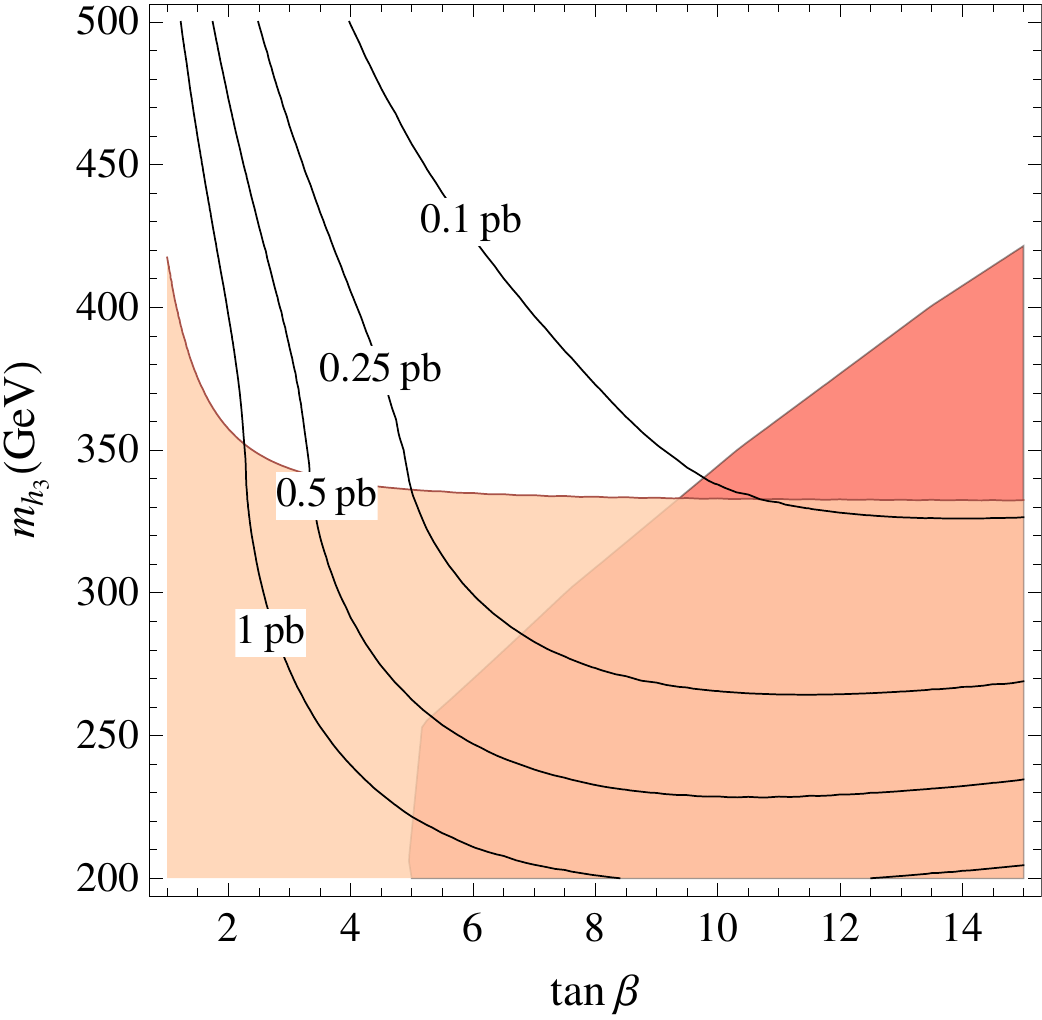}\hfill
\includegraphics[width=.48\textwidth]{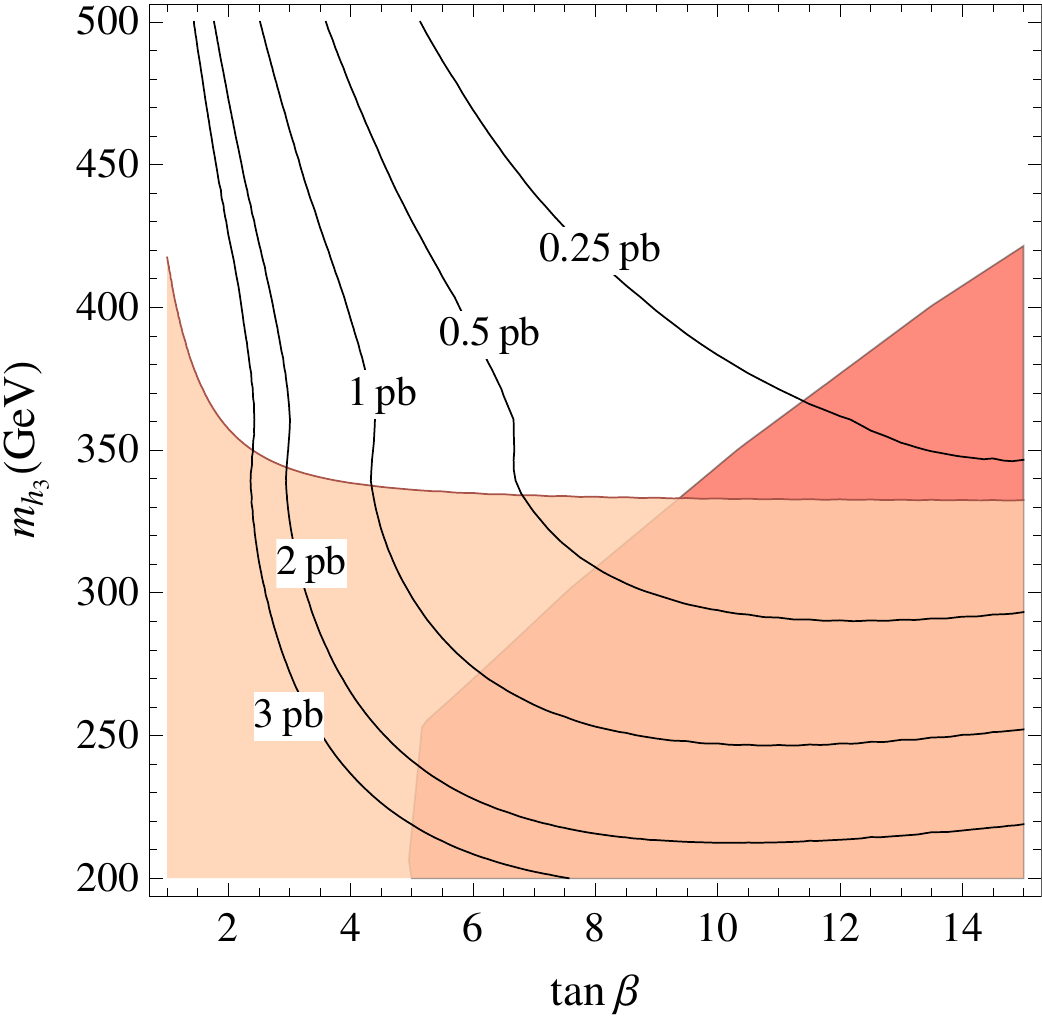}
\caption{\label{fig:MSSM-Xsec}\small MSSM. Isolines of gluon fusion production cross section $\sigma(gg\to h_3)$. Light colored region is excluded at 95$\%$C.L., the red region is excluded by CMS direct searches for $A,H\to \tau^+\tau^-$. Left: LHC8. Right: LHC14.}
\end{center}
\end{figure}

\begin{figure}
\begin{center}
\includegraphics[width=.48\textwidth]{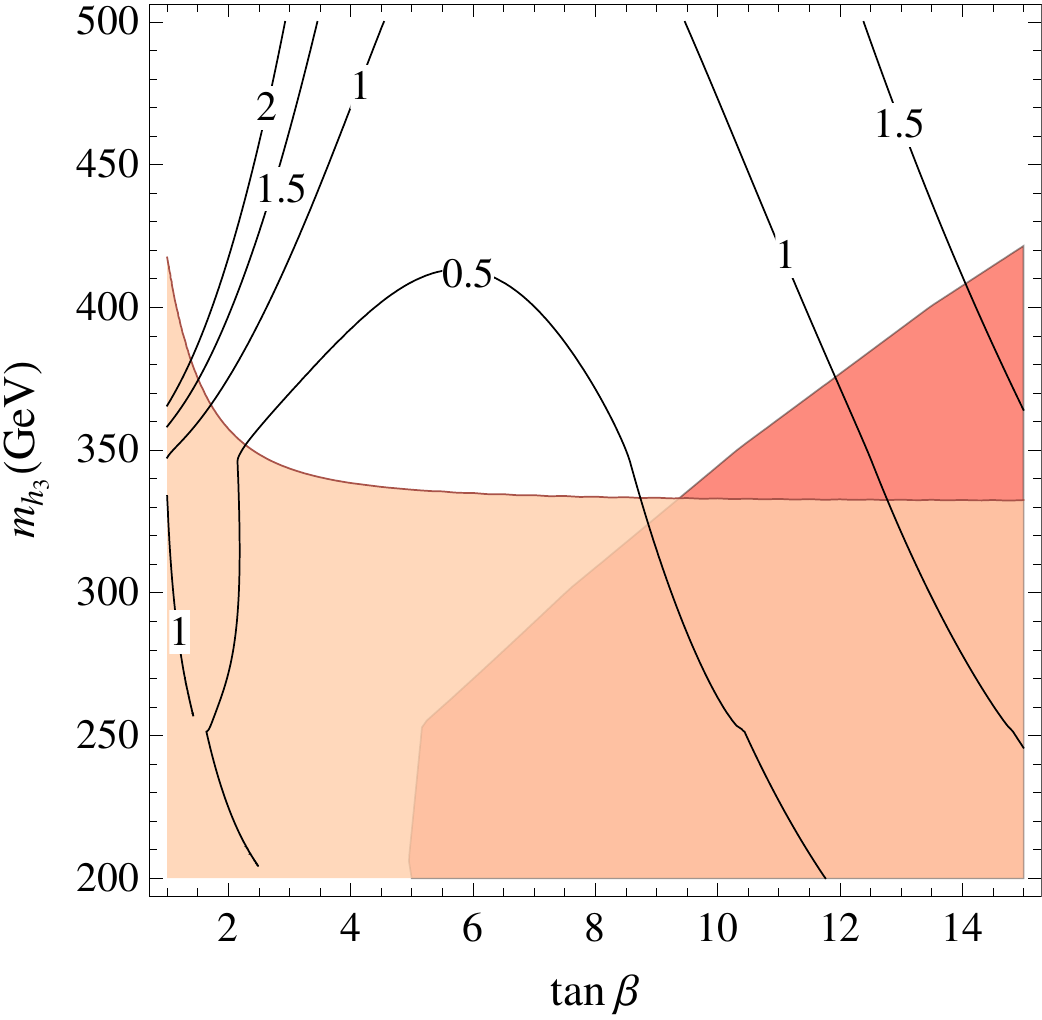}\hfill
\includegraphics[width=.48\textwidth]{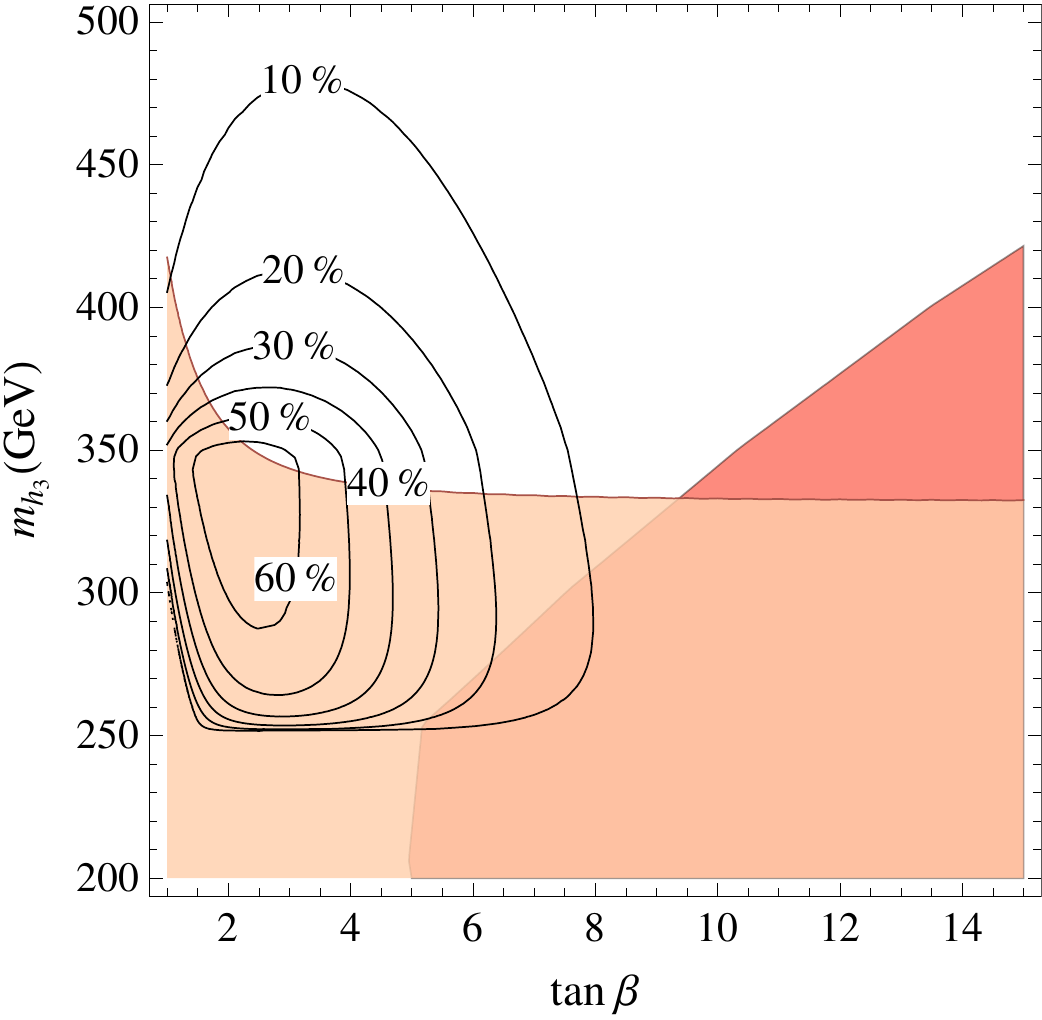}
\caption{\label{fig:MSSM-BRs}\small MSSM. Left: Isolines of the total width $\Gamma_{h_3}$ (GeV). Right: Isolines of BR$(h_3\to h h)$. The light colored region is excluded at 95\%C.L., the red region is excluded by CMS direct searches.}
\end{center}
\end{figure}

\begin{figure}
\begin{center}
\includegraphics[width=.48\textwidth]{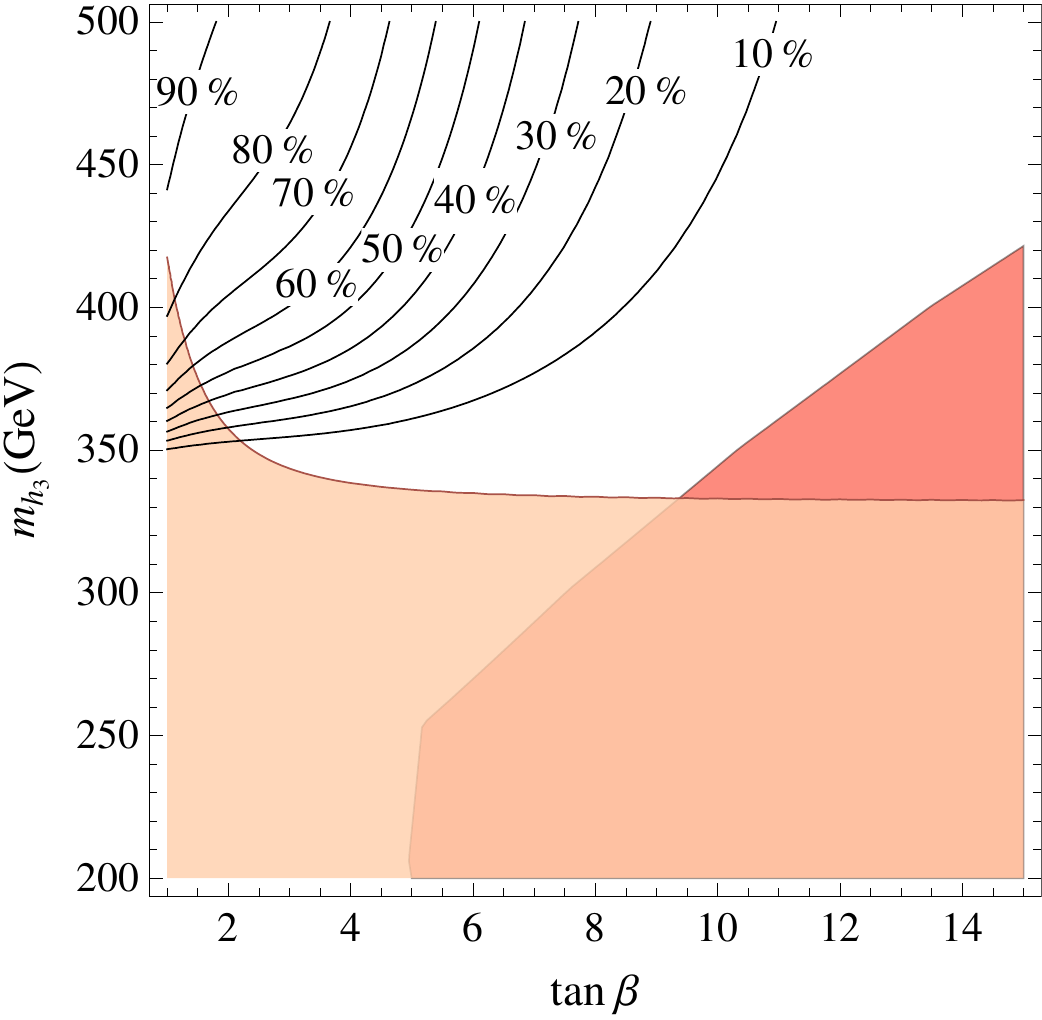}\hfill
\includegraphics[width=.48\textwidth]{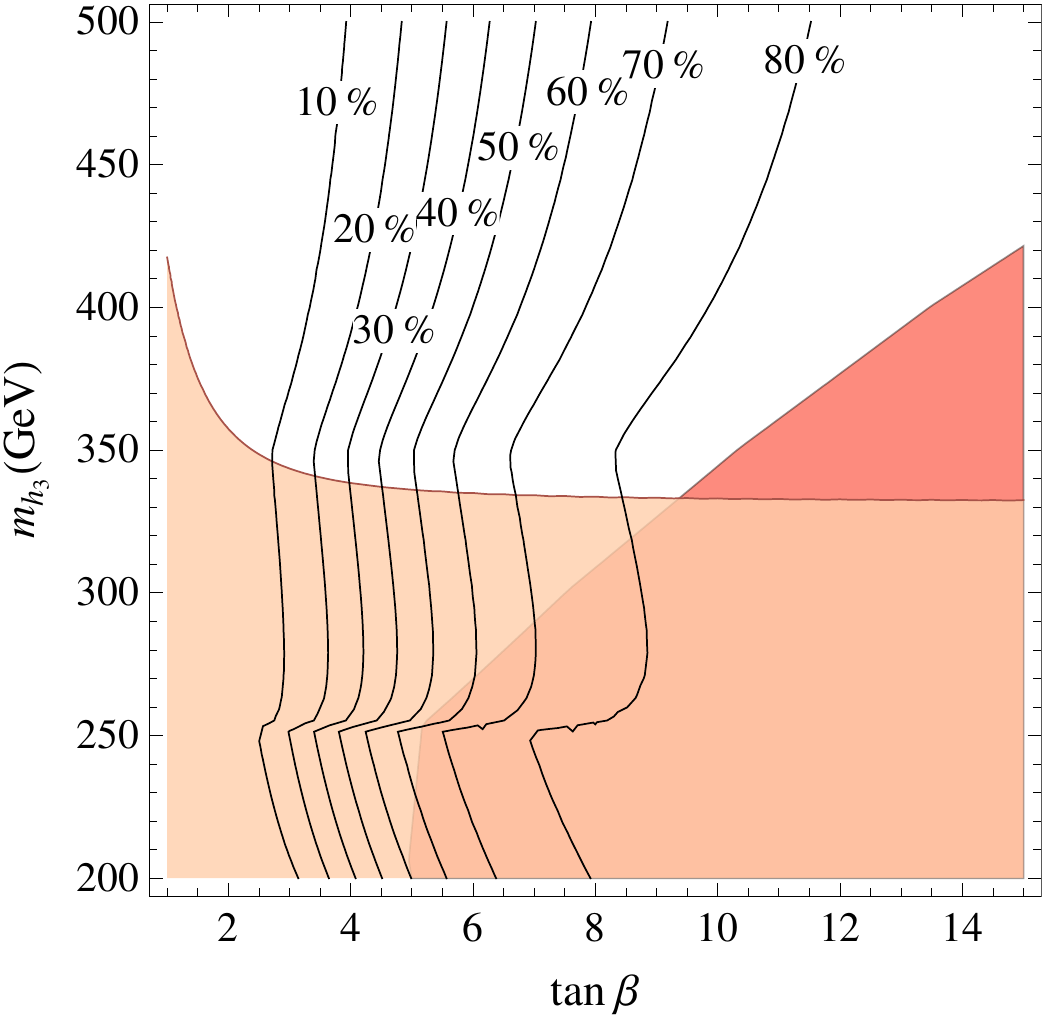}
\caption{\label{fig:MSSM-BRf}\small MSSM. Left: isolines of BR$(h_3\to t\bar t)$. Right: Isolines of BR$(h_3\to b \bar{b})$. The light colored region is excluded at 95$\%$C.L., the red region is excluded by CMS direct searches.}
\end{center}
\end{figure}

A first important difference of the MSSM versus the NMSSM is in a minimum value of $\Delta_t \gtrsim 85$ GeV that is needed to accommodate the 126 GeV Higgs boson as the lightest CP-even neutral scalar. Also for this reason, and because we have one parameter less than in the previous section, we let $\Delta_t$ vary. As a consequence, in analogy with Figure~\ref{fig:mSdecoupled-1}, we show in Figure~\ref{fig:MSSM-1} the regions allowed by current experimental data on the signal strengths 
of $h_1 = h_{\text{LHC}}$. Note that in the plane $(\tan{\beta}, m_{h_3})$ the isolines of  $\Delta_t$ are increasingly  large at lower $\tan{\beta}$: a sign of increasing fine tuning.
Finally, in analogy with Figures~\ref{fig:mSdecoupled-Xsec}-\ref{fig:mSdecoupled-BRf}, we show in Figures~\ref{fig:MSSM-Xsec}-\ref{fig:MSSM-BRf} the gluon-fusion production cross sections and the widths of $h_3$ for the MSSM case. For the production cross sections we have adopted the same procedure of the Singlet decoupled case, and performed a further check of our results with the ones recently presented in \cite{Arbey:2013jla} and \cite{Djouadi:2013vqa}, finding a very good agreement.


In the MSSM $m_A^2$ in (\ref{mA_mh}) at $\lambda=0$ is the squared mass of the neutral pseudoscalar $A$,  unlike the case of the general NMSSM, where $m_A^2$ in (\ref{mA_mh}) is only an auxiliary quantity.
In the same  $(\tan{\beta}, m_{h_3})$ plane $\sigma (gg\rightarrow A)$ is therefore also determined, which allows to delimit the currently excluded region by the direct searches for $A, h_3\rightarrow \tau^+ \tau^-$. Such a region is known to be significant, especially for growing $\tan{\beta}$. In Figures~\ref{fig:MSSM-1}-\ref{fig:MSSM-BRf} we draw the region excluded by such search, as inferred from \cite{CMS-PAS-HIG-12-050}.

\section{The NMSSM at $\lambda\gtrsim 1$ and gauge coupling unification}
\label{sec6}

As said in the Introduction we are particularly interested in the NMSSM at $\lambda$ close to one and moderate $\tan{\beta}$ to limit the fine tuning. At least in the $H$ decoupled case we have seen in Section~\ref{sec3} that this is consistent with current data. On the other hand a well known objection to the NMSSM at $\lambda \gtrsim 1$ is its compatibility with  gauge coupling unification. Requiring $\lambda$ to stay semi-perturbative up to the grand unified theory (GUT) scale bounds $\lambda$ at the weak scale at  about $0.7$ \cite{Espinosa:1991gr}. This value is in fact influenced by the presence of vector-like matter in full $SU(5)$ multiplets that slows down the running of $\lambda$  by increasing  the gauge couplings at high energies. However, even adding three vector-like five-plets at 1 TeV, in which case $\alpha_G$ still remains perturbative, does not allow $\lambda$ at the weak scale to go above $0.8$ \cite{Masip:1998jc,Barbieri:2007tu}.

There are several ways \cite{Harnik:2003rs,Chang:2004db,Birkedal:2004zx,Delgado:2005fq,Gherghetta:2011wc,Craig:2011ev,Csaki:2011xn,Hardy:2012ef} in which $\lambda$ could go to $1\div 1.5$  without spoiling unification nor affecting  the  consequences at the weak scale of the NMSSM Lagrangian, as treated above. One further possibility also makes use of two vector-like five-plets as follows. For ease of exposition let us call them  $F_{u,d} + \bar{F}_{u,d}$, where $F_u$ is a $5$ and $F_d$ a $\bar{5}$, thus containing one $SU(2)$ doublet each, $h_u$ and $h_d$, with the same quantum numbers of the standard $H_u, H_d$ used so far. Correspondingly $\bar{F}_{u,d}$ contain two doublets that we call $\bar{h}_{u,d}$. Needless to say all these are superfields. Let us further assume that the superpotential is such that:
\begin{itemize}
\item The five-plets interact with a singlet $S$ and pick up $SU(5)$-invariant masses consistently with a Peccei-Quinn symmetry;
\item The standard doublets $H_u, H_d$ mix by mass terms with $h_u$ and $h_d$, still maintaining the Peccei-Quinn symmetry, and do not interact directly with $S$.
\end{itemize}
The corresponding superpotential is 
\begin{equation}
f = \lambda_S S F_u F_d + M_u F_u \bar{F}_u + M_d F_d \bar{F}_d + m_u H_u \bar{h}_u + m_d H_d \bar{h}_d + \lambda_t H_u Q t,
\label{f}
\end{equation}
where we have also made explicit the  Yukawa coupling of the top to $H_u$.
Below these masses, all taken to be comparable, this $f$ term leaves three massless supermultiplets: 
\begin{equation}
S,~~~ \hat{H}_u = c_u H_u + s_u h_u,~~~\hat{H}_d = c_d H_d + s_d h_d,
\end{equation}
which interact through the superpotential
\begin{equation}
\hat{f}= \hat{\lambda} S \hat{H}_u \hat{H}_d + \hat{\lambda}_t \hat{H}_u Q t, ~~~\hat{\lambda} = \lambda_S s_u s_d,~~~~\hat{\lambda}_t=\lambda_t c_u .
\end{equation}
This superpotential, completed by Peccei-Quinn symmetry breaking terms at the Fermi scale, defines the effective NMSSM as discussed so far. To cure the growth of $\hat{\lambda}$ at increasing energies, the masses in (\ref{f}) will have to be crossed while $\hat{\lambda}$ is still semi-perturbative. For $\hat{\lambda} =1\div 1.5$ these masses are above 1000 TeV.

 At greater energies the running of the gauge couplings is affected, compared to the standard supersymmetric case,  by the supermultiplet $H_u$ with a top Yukawa coupling increased by a factor $1/c_u$ and by the  degenerate complete $SU(5)$ multiplets $F_{u,d} + \bar{F}_{u,d}$. To avoid a Landau pole before the GUT scale in the top Yukawa coupling  $c_u$ has to be bigger than about $1/\sqrt{2}$. As to the effect of $F_{u,d} + \bar{F}_{u,d}$, they do not alter the relative one loop running of the gauge couplings but might give rise to an exceeding growth of all of them before $M_{\text{GUT}}$ due to the presence of the coupling $\lambda_S$, which at some point will get strong. To avoid this a change of regime in the $SU(5)$-symmetric sector will have to intervene to keep under control the anomalous dimensions of the $F_{u,d}, \bar{F}_{u,d}$ superfields.

\section{Summary and conclusions}

Is the newly found Higgs boson at LHC alone or is it a member of a family of scalar particles? Now that we know that one Higgs boson exists this is a compelling question, both {\it per se} and in many motivated  different contexts. 
Here we have focussed on the case implied by the supersymmetric extensions of the SM, where naturalness requires at least one extra doublet of Higgs particles at the Fermi scale. Specifically, a minimally fine tuned case appears to be the Higgs system of the NMSSM, which also includes an extra singlet state. It is in fact not inconceivable that part of such extended Higgs system be the lightest fragments of the entire new particle spectrum of the NMSSM, with the possible exception of the LSP. It is of particular interest to know which impact the measurements -- current and foreseen -- of the different signal strengths of the newly found resonance have on this problem and how they compare with the potential of the direct 
searches of the extra states.

To have a simple characterization of the properties of the extended Higgs system we have focussed on relations between physical parameters. We have also not included any radiative effects of superpartners  other than the top-stop loop corrections to the quartic Higgs couplings, \eqref{delta-t}, which mostly affect the properties of the lightest CP even state that we identify with $h_{\rm LHC}$. While providing at least a useful reference case, we think that this is motivated by the consideration of superpartner masses at their ``naturalness limit''. Furthermore we have considered separately two limiting cases of the NMSSM, in which one CP even scalar is decoupled: either the singlet or the doublet component, $H=-s_\beta H_d + c_\beta H_u$. This allows to work in general with four effective parameters: $\lambda, \tan{\beta}$, $\Delta_t$ in (\ref{delta-t}) and the mass of the intermediate CP-even scalar, $h_2$ or $h_3$ in the two cases. The mass of $h_1=h_{\rm LHC}$ sets one relation among these parameters. A 
further relation exists in the singlet-decoupled case, of which the MSSM is a limit ($\lambda=0$).
We have not been sticking to a particular NMSSM, which might imply specific constraints on the physical parameters that we consider, but we have assumed to live in the case of negligible CP violation in the Higgs sector. 

Within this framework, even though the signal strengths of $h_{\rm LHC}$ are close to those expected in the SM, they still allow for a new further state nearby, unlike in the case of the MSSM, where a CP- even scalar heavier than $h_{\mathrm{LHC}}$ and below about 300 GeV is unlikely \cite{DAgnolo:2012mj}. This is true in both the limiting cases that we have considered, as visible in Figures~\ref{fig:mAdecoupled-1} and \ref{fig:mSdecoupled-1}, to be contrasted with Figure~\ref{fig:MSSM-1}. On the other hand, the same Figures show that the measured signal strengths of $h_{\rm LHC}$ do limit the possible values of $\lambda$, at least for moderate values of $\tan{\beta}$, which is the region mostly motivated by naturalness: $\lambda \approx 1$ is still largely allowed in the $H$-decoupled case, whereas it is borderline in the singlet decoupled situation. As commented upon in Section~\ref{sec6}, we think that $\lambda\gtrsim 1$ can be compatible with gauge coupling unification.

Most importantly from the point of view of the direct searches,\footnote{A first attempt at studying heavier Higgs decays in the NMSSM with $\lambda > 1$ was made in \cite{Cavicchia:2007dp}.} this feature reflects itself in the behavior of the new states, which quite different in the two cases, especially in their decay properties. The state $h_2$ of the $H$-decoupled case has a large BR into a pair of $h_1$, whenever it is allowed by phase space, with $VV$ as subdominant decay (Figures~\ref{fig:mAdecoupled-hh}-\ref{fig:mAdecoupled-WW}). With the production cross sections shown in Figure~\ref{fig:mAdecoupled-Xsec} its direct search at LHC8 or LHC14 may be challenging, although perhaps not impossible \cite{Gouzevitch:2013qca,Chatrchyan:2013yoa}. On the other hand the reduced value of $\lambda$ allowed in the singlet decoupled case makes the $b\bar{b}$ channel, and so the $\tau\bar{\tau}$, most important, below the $t\bar{t}$ threshold (Figures~\ref{fig:mSdecoupled-BRs} and \ref{fig:mSdecoupled-BRf}).  This makes the state $h_3$ relatively more similar to the CP-even $H$ state of the MSSM (Figures~\ref{fig:MSSM-BRs} and \ref{fig:MSSM-BRf}), 
which is being 
actively searched. 
An important point in this comparison is that in the  NMSSM there are two CP-odd states and that the lightest of the two may be quite different from the state $A$ of the MSSM unless in a special $S$-decoupled case.  

It will be interesting to follow the progression of the searches of the Higgs system of the NMSSM, directly or indirectly through the more precise measurements of the $h_1$ properties. We believe that the framework outlined here should allow one to systematize these searches in a clear way. We also think that they should be pursued actively and independently from the searches of the superpartners.

\subsubsection*{Acknowledgments}
We thank Raffaele Tito D'Agnolo, Marco Farina, Alessandro Strumia and Caterina Vernieri for useful discussions.
This work is supported in part by the European Programme ``Unification in the LHC Era",  contract PITN-GA-2009-237920 (UNILHC), MIUR under contract 2010YJ2NYW-010, the ESF grants 8943, MJD140 and MTT8, by the recurrent financing SF0690030s09 project and by the European Union through the European Regional Development Fund.

\bibliographystyle{My}
\small
\bibliography{NMSSM_arxiv_v2}

\providecommand{\href}[2]{#2}\begingroup\raggedright\begin{thebibliography}{10}

\bibitem{Dimopoulos:1995mi}
S.~Dimopoulos and G.~Giudice,
  \href{http://dx.doi.org/10.1016/0370-2693(95)00961-J}{{\em Phys.Lett.} {\bf
  B357} (1995)  573--578},
\href{http://arxiv.org/abs/hep-ph/9507282}{{\tt arXiv:hep-ph/9507282
  [hep-ph]}}.

\bibitem{Cohen:1996vb}
A.~G. Cohen, D.~Kaplan, and A.~Nelson,
  \href{http://dx.doi.org/10.1016/S0370-2693(96)01183-5}{{\em Phys.Lett.} {\bf
  B388} (1996)  588--598},
\href{http://arxiv.org/abs/hep-ph/9607394}{{\tt arXiv:hep-ph/9607394
  [hep-ph]}}.

\bibitem{Barbieri:2009ev}
R.~Barbieri and D.~Pappadopulo,
  \href{http://dx.doi.org/10.1088/1126-6708/2009/10/061}{{\em JHEP} {\bf 0910}
  (2009)  061},
\href{http://arxiv.org/abs/0906.4546}{{\tt arXiv:0906.4546 [hep-ph]}}.

\bibitem{Papucci:2011wy}
M.~Papucci, J.~T. Ruderman, and A.~Weiler,
  \href{http://dx.doi.org/10.1007/JHEP09(2012)035}{{\em JHEP} {\bf 1209} (2012)
   035},
\href{http://arxiv.org/abs/1110.6926}{{\tt arXiv:1110.6926 [hep-ph]}}.

\bibitem{Aad:2012tfa}
{\bf ATLAS Collaboration}, G.~Aad {\em et al.,}
  \href{http://dx.doi.org/10.1016/j.physletb.2012.08.020}{{\em Phys.Lett.} {\bf
  B716} (2012)  1--29},
\href{http://arxiv.org/abs/1207.7214}{{\tt arXiv:1207.7214 [hep-ex]}}.

\bibitem{Chatrchyan:2012ufa}
{\bf CMS Collaboration}, S.~Chatrchyan {\em et al.,}
  \href{http://dx.doi.org/10.1016/j.physletb.2012.08.021}{{\em Phys.Lett.} {\bf
  B716} (2012)  30--61},
\href{http://arxiv.org/abs/1207.7235}{{\tt arXiv:1207.7235 [hep-ex]}}.

\bibitem{ATLAS}
 {\bf ATLAS Collaboration}, F. Hubaut. Talk at the Moriond 2013 EW
  session.\\[0.15cm] {\bf ATLAS Collaboration}, E. Mountricha. Talk at the
  Moriond 2013 QCD session.\\[0.15cm] {\bf ATLAS Collaboration}, V. Martin.
  Talk at the Moriond 2013 EW session.\\[0.15cm] {\bf ATLAS Collaboration},
  ATLAS-CONF-2013-009.\\[0.15cm] {\bf ATLAS Collaboration},
  ATLAS-CONF-2013-010.\\[0.15cm] {\bf ATLAS Collaboration},
  ATLAS-CONF-2013-011.\\[0.15cm] {\bf ATLAS Collaboration},
  ATLAS-CONF-2013-012.\\[0.15cm] {\bf ATLAS Collaboration},
  ATLAS-CONF-2013-013.\\[0.15cm] {\bf ATLAS Collaboration},
  ATLAS-CONF-2013-014.\\[0.15cm] {\bf ATLAS Collaboration},
  ATLAS-CONF-2013-030.

\bibitem{CMS}
 {\bf CMS Collaboration}, G. Gomez-Ceballos. Talk at the Moriond 2013 EW
  session.\\[0.15cm] {\bf CMS Collaboration}, M. Shen. Talk at the Moriond 2013
  QCD session.\\[0.15cm] {\bf CMS Collaboration}, B. Mansoulie. Talk at the
  Moriond 2013 EW session.\\[0.15cm] {\bf CMS Collaboration}, V. Dutta. Talk at
  the Moriond 2013 EW session.\\[0.15cm] {\bf CMS Collaboration},
  CMS-PAS-HIG-12-050.\\[0.15cm] {\bf CMS Collaboration},
  CMS-PAS-HIG-13-001.\\[0.15cm] {\bf CMS Collaboration},
  CMS-PAS-HIG-13-002.\\[0.15cm] {\bf CMS Collaboration},
  CMS-PAS-HIG-13-003.\\[0.15cm] {\bf CMS Collaboration},
  CMS-PAS-HIG-13-004.\\[0.15cm] {\bf CMS Collaboration},
  CMS-PAS-HIG-13-006.\\[0.15cm] {\bf CMS Collaboration}, CMS-PAS-HIG-13-009.

\bibitem{tevatron:2013}
{\bf CDF and D0 Collaborations}, L.~\v{Z}ivkovi\'{c}. Talk at the Moriond 2013
  EW session.

\bibitem{Fayet:1974pd}
P.~Fayet
\href{http://dx.doi.org/10.1016/0550-3213(75)90636-7}{{\em Nucl.Phys.} {\bf
  B90} (1975)  104--124}.

\bibitem{Ellwanger:2009dp}
U.~Ellwanger, C.~Hugonie, and A.~M. Teixeira,
  \href{http://dx.doi.org/10.1016/j.physrep.2010.07.001}{{\em Phys.Rept.} {\bf
  496} (2010)  1--77},
\href{http://arxiv.org/abs/0910.1785}{{\tt arXiv:0910.1785 [hep-ph]}}.

\bibitem{Barbieri:2006bg}
R.~Barbieri, L.~J. Hall, Y.~Nomura, and V.~S. Rychkov,
  \href{http://dx.doi.org/10.1103/PhysRevD.75.035007}{{\em Phys.Rev.} {\bf D75}
  (2007)  035007},
\href{http://arxiv.org/abs/hep-ph/0607332}{{\tt arXiv:hep-ph/0607332
  [hep-ph]}}.

\bibitem{Hall:2011aa}
L.~J. Hall, D.~Pinner, and J.~T. Ruderman,
  \href{http://dx.doi.org/10.1007/JHEP04(2012)131}{{\em JHEP} {\bf 1204} (2012)
   131},
\href{http://arxiv.org/abs/1112.2703}{{\tt arXiv:1112.2703 [hep-ph]}}.

\bibitem{Agashe:2012zq}
K.~Agashe, Y.~Cui, and R.~Franceschini,
  \href{http://dx.doi.org/10.1007/JHEP02(2013)031}{{\em JHEP} {\bf 1302} (2013)
   031},
\href{http://arxiv.org/abs/1209.2115}{{\tt arXiv:1209.2115 [hep-ph]}}.

\bibitem{Gherghetta:2012gb}
T.~Gherghetta, B.~von Harling, A.~D. Medina, and M.~A. Schmidt,
  \href{http://dx.doi.org/10.1007/JHEP02(2013)032}{{\em JHEP} {\bf 1302} (2013)
   032},
\href{http://arxiv.org/abs/1212.5243}{{\tt arXiv:1212.5243 [hep-ph]}}.

\bibitem{Gupta:2012fy}
R.~S. Gupta, M.~Montull, and F.~Riva,
\href{http://arxiv.org/abs/1212.5240}{{\tt arXiv:1212.5240 [hep-ph]}}.

\bibitem{DAgnolo:2012mj}
R.~T. D'Agnolo, E.~Kuflik, and M.~Zanetti,
\href{http://arxiv.org/abs/1212.1165}{{\tt arXiv:1212.1165 [hep-ph]}}.

\bibitem{Djouadi:2005gj}
A.~Djouadi \href{http://dx.doi.org/10.1016/j.physrep.2007.10.005}{{\em
  Phys.Rept.} {\bf 459} (2008)  1--241},
\href{http://arxiv.org/abs/hep-ph/0503173}{{\tt arXiv:hep-ph/0503173
  [hep-ph]}}.

\bibitem{King:2012tr}
S.~F. King, M.~Muhlleitner, R.~Nevzorov, and K.~Walz,
  \href{http://dx.doi.org/10.1016/j.nuclphysb.2013.01.020}{{\em Nucl.Phys.}
  {\bf B870} (2013)  323--352},
\href{http://arxiv.org/abs/1211.5074}{{\tt arXiv:1211.5074 [hep-ph]}}.

\bibitem{Cheung:2013bn}
C.~Cheung, S.~D. McDermott, and K.~M. Zurek,
\href{http://arxiv.org/abs/1302.0314}{{\tt arXiv:1302.0314 [hep-ph]}}.

\bibitem{Choi:2012he}
K.~Choi, S.~H. Im, K.~S. Jeong, and M.~Yamaguchi,
  \href{http://dx.doi.org/10.1007/JHEP02(2013)090}{{\em JHEP} {\bf 1302} (2013)
   090},
\href{http://arxiv.org/abs/1211.0875}{{\tt arXiv:1211.0875 [hep-ph]}}.

\bibitem{Giardino:2013bma}
P.~P. Giardino, K.~Kannike, I.~Masina, M.~Raidal, and A.~Strumia,
\href{http://arxiv.org/abs/1303.3570}{{\tt arXiv:1303.3570 [hep-ph]}}.

\bibitem{Carena:1995bx}
M.~S. Carena, J.~Espinosa, M.~Quiros, and C.~Wagner,
  \href{http://dx.doi.org/10.1016/0370-2693(95)00694-G}{{\em Phys.Lett.} {\bf
  B355} (1995)  209--221},
\href{http://arxiv.org/abs/hep-ph/9504316}{{\tt arXiv:hep-ph/9504316
  [hep-ph]}}.

\bibitem{Dittmaier:2011ti}
{\bf LHC Higgs Cross Section Working Group}, S.~Dittmaier {\em et al.,}
  \href{http://arxiv.org/abs/1101.0593}{{\tt arXiv:1101.0593 [hep-ph]}}.
Updates on
  \href{https://twiki.cern.ch/twiki/bin/view/LHCPhysics/CrossSections}{https:/%
/twiki.cern.ch/twiki/bin/view/LHCPhysics/CrossSections}.

\bibitem{Baglio:2012np}
J.~Baglio, A.~Djouadi, R.~Grober, M.~Muhlleitner, J.~Quevillon, {\em et al.,}
\href{http://arxiv.org/abs/1212.5581}{{\tt arXiv:1212.5581 [hep-ph]}}.

\bibitem{Misiak:2006zs}
M.~Misiak, H.~Asatrian, K.~Bieri, M.~Czakon, A.~Czarnecki, {\em et al.,}
  \href{http://dx.doi.org/10.1103/PhysRevLett.98.022002}{{\em Phys.Rev.Lett.}
  {\bf 98} (2007)  022002},
\href{http://arxiv.org/abs/hep-ph/0609232}{{\tt arXiv:hep-ph/0609232
  [hep-ph]}}.

\bibitem{Azatov:2012qz}
A.~Azatov and J.~Galloway,
  \href{http://dx.doi.org/10.1142/S0217751X13300044}{{\em Int.J.Mod.Phys.} {\bf
  A28} (2013)  1330004},
\href{http://arxiv.org/abs/1212.1380}{{\tt arXiv:1212.1380 [hep-ph]}}.

\bibitem{Carmi:2012in}
D.~Carmi, A.~Falkowski, E.~Kuflik, T.~Volansky, and J.~Zupan,
  \href{http://dx.doi.org/10.1007/JHEP10(2012)196}{{\em JHEP} {\bf 1210} (2012)
   196},
\href{http://arxiv.org/abs/1207.1718}{{\tt arXiv:1207.1718 [hep-ph]}}.

\bibitem{Anastasiou:2009kn}
C.~Anastasiou, S.~Bucherer, and Z.~Kunszt,
  \href{http://dx.doi.org/10.1088/1126-6708/2009/10/068}{{\em JHEP} {\bf 0910}
  (2009)  068},
\href{http://arxiv.org/abs/0907.2362}{{\tt arXiv:0907.2362 [hep-ph]}}.

\bibitem{Spira:1995rr}
M.~Spira, A.~Djouadi, D.~Graudenz, and P.~Zerwas,
  \href{http://dx.doi.org/10.1016/0550-3213(95)00379-7}{{\em Nucl.Phys.} {\bf
  B453} (1995)  17--82},
\href{http://arxiv.org/abs/hep-ph/9504378}{{\tt arXiv:hep-ph/9504378
  [hep-ph]}}.

\bibitem{Spira:1995mt}
M.~Spira
\href{http://arxiv.org/abs/hep-ph/9510347}{{\tt arXiv:hep-ph/9510347
  [hep-ph]}}.

\bibitem{Djouadi:2013vqa}
A.~Djouadi and J.~Quevillon,
\href{http://arxiv.org/abs/1304.1787}{{\tt arXiv:1304.1787 [hep-ph]}}.

\bibitem{Maiani:2012ij}
L.~Maiani, A.~Polosa, and V.~Riquer,
  \href{http://dx.doi.org/10.1088/1367-2630/14/7/073029}{{\em New J.Phys.} {\bf
  14} (2012)  073029},
\href{http://arxiv.org/abs/1202.5998}{{\tt arXiv:1202.5998 [hep-ph]}}.

\bibitem{Maiani:2012qf}
L.~Maiani, A.~Polosa, and V.~Riquer,
  \href{http://dx.doi.org/10.1016/j.physletb.2012.10.041}{{\em Phys.Lett.} {\bf
  B718} (2012)  465--468},
\href{http://arxiv.org/abs/1209.4816}{{\tt arXiv:1209.4816 [hep-ph]}}.

\bibitem{Arbey:2013jla}
A.~Arbey, M.~Battaglia, and F.~Mahmoudi,
\href{http://arxiv.org/abs/1303.7450}{{\tt arXiv:1303.7450 [hep-ph]}}.

\bibitem{CMS-PAS-HIG-12-050}
{\bf {CMS Collaboration}}. CMS-PAS-HIG-12-050.

\bibitem{Espinosa:1991gr}
J.~Espinosa and M.~Quiros,
\href{http://dx.doi.org/10.1016/0370-2693(92)91846-2}{{\em Phys.Lett.} {\bf
  B279} (1992)  92--97}.

\bibitem{Masip:1998jc}
M.~Masip, R.~Munoz-Tapia, and A.~Pomarol,
  \href{http://dx.doi.org/10.1103/PhysRevD.57.5340}{{\em Phys.Rev.} {\bf D57}
  (1998)  R5340},
\href{http://arxiv.org/abs/hep-ph/9801437}{{\tt arXiv:hep-ph/9801437
  [hep-ph]}}.

\bibitem{Barbieri:2007tu}
R.~Barbieri, L.~J. Hall, A.~Y. Papaioannou, D.~Pappadopulo, and V.~S. Rychkov,
  \href{http://dx.doi.org/10.1088/1126-6708/2008/03/005}{{\em JHEP} {\bf 0803}
  (2008)  005},
\href{http://arxiv.org/abs/0712.2903}{{\tt arXiv:0712.2903 [hep-ph]}}.

\bibitem{Harnik:2003rs}
R.~Harnik, G.~D. Kribs, D.~T. Larson, and H.~Murayama,
  \href{http://dx.doi.org/10.1103/PhysRevD.70.015002}{{\em Phys.Rev.} {\bf D70}
  (2004)  015002},
\href{http://arxiv.org/abs/hep-ph/0311349}{{\tt arXiv:hep-ph/0311349
  [hep-ph]}}.

\bibitem{Chang:2004db}
S.~Chang, C.~Kilic, and R.~Mahbubani,
  \href{http://dx.doi.org/10.1103/PhysRevD.71.015003}{{\em Phys.Rev.} {\bf D71}
  (2005)  015003},
\href{http://arxiv.org/abs/hep-ph/0405267}{{\tt arXiv:hep-ph/0405267
  [hep-ph]}}.

\bibitem{Birkedal:2004zx}
A.~Birkedal, Z.~Chacko, and Y.~Nomura,
  \href{http://dx.doi.org/10.1103/PhysRevD.71.015006}{{\em Phys.Rev.} {\bf D71}
  (2005)  015006},
\href{http://arxiv.org/abs/hep-ph/0408329}{{\tt arXiv:hep-ph/0408329
  [hep-ph]}}.

\bibitem{Delgado:2005fq}
A.~Delgado and T.~M. Tait,
  \href{http://dx.doi.org/10.1088/1126-6708/2005/07/023}{{\em JHEP} {\bf 0507}
  (2005)  023},
\href{http://arxiv.org/abs/hep-ph/0504224}{{\tt arXiv:hep-ph/0504224
  [hep-ph]}}.

\bibitem{Gherghetta:2011wc}
T.~Gherghetta, B.~von Harling, and N.~Setzer,
  \href{http://dx.doi.org/10.1007/JHEP07(2011)011}{{\em JHEP} {\bf 1107} (2011)
   011},
\href{http://arxiv.org/abs/1104.3171}{{\tt arXiv:1104.3171 [hep-ph]}}.

\bibitem{Craig:2011ev}
N.~Craig, D.~Stolarski, and J.~Thaler,
  \href{http://dx.doi.org/10.1007/JHEP11(2011)145}{{\em JHEP} {\bf 1111} (2011)
   145},
\href{http://arxiv.org/abs/1106.2164}{{\tt arXiv:1106.2164 [hep-ph]}}.

\bibitem{Csaki:2011xn}
C.~Csaki, Y.~Shirman, and J.~Terning,
  \href{http://dx.doi.org/10.1103/PhysRevD.84.095011}{{\em Phys.Rev.} {\bf D84}
  (2011)  095011},
\href{http://arxiv.org/abs/1106.3074}{{\tt arXiv:1106.3074 [hep-ph]}}.

\bibitem{Hardy:2012ef}
E.~Hardy, J.~March-Russell, and J.~Unwin,
  \href{http://dx.doi.org/10.1007/JHEP10(2012)072}{{\em JHEP} {\bf 1210} (2012)
   072},
\href{http://arxiv.org/abs/1207.1435}{{\tt arXiv:1207.1435 [hep-ph]}}.

\bibitem{Cavicchia:2007dp}
L.~Cavicchia, R.~Franceschini, and V.~S. Rychkov,
  \href{http://dx.doi.org/10.1103/PhysRevD.77.055006}{{\em Phys.Rev.} {\bf D77}
  (2008)  055006},
\href{http://arxiv.org/abs/0710.5750}{{\tt arXiv:0710.5750 [hep-ph]}}.

\bibitem{Gouzevitch:2013qca}
M.~Gouzevitch, A.~Oliveira, J.~Rojo, R.~Rosenfeld, G.~Salam, {\em et al.,}
\href{http://arxiv.org/abs/1303.6636}{{\tt arXiv:1303.6636 [hep-ph]}}.

\bibitem{Chatrchyan:2013yoa}
{\bf CMS Collaboration}, S.~Chatrchyan {\em et al.,}
\href{http://arxiv.org/abs/1304.0213}{{\tt arXiv:1304.0213 [hep-ex]}}.

\end{thebibliography}\endgroup

%
\end{document}